\documentclass[aps,pre,reprint,superscriptaddress,nofootinbib,floatfix,longbibliography]{revtex4-2}

\usepackage[T1]{fontenc}
\usepackage[utf8]{inputenc}
\usepackage{lmodern}
\usepackage{amsmath,amssymb,amsfonts,bm,mathtools,mathrsfs}
\usepackage{graphicx}
\usepackage{booktabs}
\usepackage{siunitx}
\usepackage{braket}
\usepackage{microtype}
\usepackage{xcolor}
\usepackage[colorlinks=true,linkcolor=blue,citecolor=blue,urlcolor=blue]{hyperref}
\usepackage{orcidlink}

\newcommand{\Tr}{\operatorname{Tr}}
\newcommand{\Ne}{\hat N_{\mathrm e}}
\newcommand{\Tb}{T_{\mathrm b}}
\newcommand{\ZD}{Z_{\mathrm D}}
\newcommand{\CD}{C_{\mathrm D}}
\newcommand{\Ueff}{U_{\mathrm{eff}}}
\newcommand{\Wu}{\mathcal W_{U(1)}}
\newcommand{\Wg}{\mathcal W_{\mathrm{glob}}}
\newcommand{\WN}{\mathcal W_N}

\newcommand{\Est}{\Delta E_B}
\newcommand{\Lmix}{\mathcal L^{\mathrm{mix}}_{U(1)}}
\newcommand{\GU}{\mathcal G_{U(1)}}
\newcommand{\WNS}{\mathcal W_{N,S}}
\newcommand{\LNS}{\mathcal L^{\mathrm{mix}}_{N,S}}
\newcommand{\EpNS}{E^{\mathrm{pass}}_{N,S}}
\newcommand{\EpU}{E^{\mathrm{pass}}_{U(1)}}

\begin{document}

\title{Interaction-enabled charging and thermal crossover of the spin-work deficit in a Hubbard double quantum dot}
\author{Khalil Loukhssami~\orcidlink{0009-0008-7934-3001}}
\email{khalil.loukhssami@um5r.ac.ma}
\affiliation{ESMaR, Department of Physics, Faculty of Sciences, Mohammed V University in Rabat, Rabat, Morocco}
\author{Cheng Shang~\orcidlink{0000-0001-8393-2329}}
\email{cheng.shang@riken.jp}
\affiliation{Analytical Quantum Complexity RIKEN Hakubi Research Team, RIKEN Center for Quantum Computing, Japan}
\author{Morad El Baz~\orcidlink{0000-0001-5201-2212}}\email{m.elbaz@um5r.ac.ma }
\affiliation{ESMaR, Department of Physics, Faculty of Sciences, Mohammed V University in Rabat, Rabat, Morocco}
\author{Khadija El Hawary~\orcidlink{0000-0003-1204-3297}}
\email{khadija_elhawary@um5.ac.ma}
\affiliation{ESMaR, Department of Physics, Faculty of Sciences, Mohammed V University in Rabat, Rabat, Morocco}

\date{November 4, 2026}
\hypersetup{pdftitle={Interaction-enabled charging and thermal crossover of the spin-work deficit in a Hubbard double quantum dot},pdfauthor={Khalil Loukhssami, Cheng Shang, Morad El Baz, Khadija El Hawary}}

\begin{abstract}
We study how conservation of total spin affects the extractable work of an
interacting Hubbard double quantum dot charged in the presence of unresolved
quasistatic detuning fluctuations. The full fermionic Fock space is used to
model preparation through electronic reservoirs, whereas the additional work
deficit associated with spin conservation arises entirely from the singlet
and triplet states at half filling. At zero detuning, modulation of the symmetric
tunnel barrier couples two singlets of even parity. The corresponding transition
matrix element vanishes when $U=V$. An exact formula for the additional work
deficit follows from passive spectral rearrangement under two extraction
constraints: conservation of particle number alone and conservation of both
particle number and total spin. At low temperature, an expansion in the detuning
response gives a leading estimate of the noise amplitude at which this deficit
first appears. Increasing the preparation temperature changes the triplet
population and produces a crossover at $s_2=q$, where the additional
spin constraint becomes nonbinding for the averaged state, even though
both ergotropies may continue to decrease. We obtain a semi-analytic
expression for this crossover by separating the thermal Gibbs weights from the
singlet dynamics under a fixed charging pulse. Averaging over symmetric detuning
disorder also restores parity in the density matrix and allows singlet eigenvalues
of different parity to exchange order, changing the branch that determines the
thermal boundary. The work values are optima over specified classes of unitary
operations and do not establish the yield of a finite discharge protocol.
\end{abstract}
\maketitle

\raggedbottom
\setlength{\parskip}{0pt}
\section{Introduction}
\label{sec:introduction}

The energy stored in a quantum system does not necessarily equal the work
that can be recovered from it. For quantum batteries, external driving prepares nonequilibrium
states, but the amount of work that can subsequently be extracted depends on the available
discharge operations. Passivity provides the basis
for this description: a passive state admits no positive work extraction
through cyclic unitary operations, while ergotropy quantifies the maximum
work recoverable under arbitrary unitary transformations
\cite{Pusz1978,Lenard1978,Allahverdyan2004}. Quantum coherence,
interactions, and collective operations can influence charging and
work extraction, as established in several quantum-battery models
\cite{Alicki2013,Campaioli2017,Andolina2019} and discussed in the
review by Campaioli \emph{et al.} \cite{Campaioli2024}.
In a physical device, the microscopic dynamics restrict the operations that can be implemented.
Conserved quantities can prohibit
population rearrangements that would otherwise lower the energy of
a charged state. Superselection rules constrain accessible quantum resources and
mechanical work \cite{Bartlett2007,Vaccaro2008}. More recently,
Mitsuhashi \emph{et al.} established criteria for passivity
under symmetry-preserving operations and characterized complete
passivity for continuous symmetries in terms of generalized Gibbs
ensembles \cite{Mitsuhashi2022}. Related work has used nonequilibrium
work statistics to probe symmetry breaking near excited-state quantum
phase transitions \cite{Mzaouali2021WorkStatistics}. Here we instead
compare extraction optima under exact charge and spin conservation in an
electrically driven fermionic quantum dot.

Semiconductor double quantum dots (DQDs) provide a microscopic setting
in which electronic interactions, charge occupation, and electrical
control can be treated within the same Hamiltonian. Gate voltages
regulate the orbital energies and interdot tunneling, while Coulomb
blockade determines the accessible charge configurations
\cite{vanDerWiel2002}. Extended Fermi--Hubbard descriptions capture
the interplay between tunneling, on-site repulsion, and interdot
Coulomb interactions that govern the charge-addition spectrum
\cite{Burkard1999,Yang2011,Wang2011}. The experimental realization of tunable
Hubbard parameters in semiconductor quantum-dot arrays by
Hensgens \emph{et al.} \cite{Hensgens2017} provides a direct
connection between this description and gate-defined devices.
For a spinful DQD with one orbital per site, the four
fermionic modes span a 16-dimensional Fock space, naturally
partitioned into sectors containing zero to four electrons.
The resulting description includes double occupation and spin degrees of freedom omitted by
effective models of charge qubits. Quantum correlations and their response
to dephasing have also been studied in coupled semiconductor quantum dots
using excitonic models \cite{AitMansour2020InAs}. Hubbard quantum-dot
studies have examined correlations and entanglement in the underlying
fermionic setting \cite{Abaach2021,Abaach2022PRA}. The single-orbital description
is nevertheless restricted to an appropriate low-energy window,
as illustrated by comparisons between Hubbard models and
configuration-interaction calculations \cite{Wang2011,Willmes2026}.

The preparation of the electronic state is particularly relevant
when different charge sectors are retained. During equilibration
with electronic reservoirs, the occupations are governed by the
addition energies, temperature, and chemical potential, and can
be described by a grand-canonical ensemble
\cite{vanDerWiel2002,Wang2011}. Once the reservoirs are disconnected,
the electron number is conserved by the usual tunneling and
electrostatic controls. The probabilities assigned to the different
charge sectors consequently remain fixed during isolated charging,
although their internal states can evolve. Thermal occupation alone
therefore does not determine whether a sector can absorb energy
from a particular driving field. Particle number also restricts extraction: an allowed unitary
cannot transfer population between sectors with different numbers of electrons. For the battery
mechanism studied below, the $N=2$ sector
contains the relevant interacting singlet--triplet structure; retaining
all sectors during loading determines its statistical weight when
initialization is not charge postselected. The preparation requires this larger space, but the
crossover itself is a property of the six states at half filling.

The choice of electrical control introduces further selection rules.
In a double quantum dot, changing the interdot tunnel barrier and
varying the orbital detuning correspond to different perturbations
of the electronic Hamiltonian. Symmetric barrier operation has
been demonstrated in semiconductor spin qubits, where tuning
the exchange interaction near a symmetric operating point reduces
sensitivity to certain charge fluctuations
\cite{Reed2016,Martins2016}. The difference between barrier and detuning control has also been examined theoretically using
Hubbard and microscopic confinement models \cite{Shim2018}.
These developments motivate barrier modulation as a control
mechanism, although the singlet transition studied here is
different from the exchange-gate operations implemented in those
experiments. Electrical fluctuations remain relevant even at
symmetric operating points. Measurements in semiconductor
quantum dots reveal substantial low-frequency charge noise,
with spectra and amplitudes that depend on the device and
experimental conditions
\cite{Dial2013,Connors2019,Connors2022,Paladino2014}.
Reduced sensitivity to detuning at first order therefore does not eliminate the influence of
disorder on the charged state or its extractable work.

Interacting quantum batteries based on Hubbard models have
previously been investigated in ultracold atomic systems, where
particle statistics, filling, and temperature influence charging
performance \cite{Konar2022}.
Related quantum-battery studies have examined mechanisms for protecting
stored work and the role of many-body structure in charging performance
\cite{Lu2025TopologicalQB,ShuklaShang2026}.
Temperature-dependent ergotropy
thresholds have also been established for dissipatively charged
spin batteries \cite{Yang2026Thermal}, whereas the present study concerns the \emph{difference
between two extraction optima} that obey different conservation laws. A recent
Hubbard--Holstein battery study highlights the distinction between
formal ergotropy and work extraction through a specified
battery--load coupling \cite{Ostberg2026}. Semiconductor quantum-dot batteries
have also been considered within effective charge-qubit models,
including studies of correlation-dependent charging and the
response of population-dominated ergotropy to detuning noise
\cite{Loukhssami2026PS,Loukhssami2026}. In those descriptions,
the relevant charge configurations are fixed at the level of
the effective qubits. A spinful Hubbard description instead
allows the preparation to populate several electron-number
sectors while electrical charging acts within each sector
according to its own spectral and symmetry properties.
This raises the question of how the thermally occupied
configurations, the transitions selected by the driving field,
and the conservation laws governing extraction jointly determine
the useful work stored in the device.

Here, we study how spin conservation limits the work extractable from a
Hubbard double quantum dot charged in the presence of unresolved quasistatic
detuning fluctuations. We compare optimal extraction under particle number conservation with
the stricter case in which total spin is conserved as well, and derive an exact spectral
expression for their difference. This difference arises solely from the
singlet--triplet sector at half filling. Symmetric barrier modulation
provides the interaction-enabled charging channel, while grand-canonical
preparation determines the charge-sector weights.

We estimate the onset of the deficit perturbatively at weak noise and examine how increasing
preparation temperature removes it by changing the triplet occupation.
The thermal weights can be separated from the singlet dynamics, which are independent of
preparation temperature, yielding a semi-analytic expression for the crossover temperature.
The parity-resolved singlet spectrum explains the change of branch along
this boundary. Throughout, the quoted work values are maxima over unitary operations satisfying
the stated conservation laws, rather than yields obtained with a specified discharge pulse.
\section{Microscopic model and charge-sector preparation}
\label{sec:model}
\subsection{Fermionic Hilbert space and charge conservation}
\label{subsec:fock}
A charge-qubit reduction fixes the accessible occupations before the
charging dynamics is specified. To examine charge preparation, we retain the full Fock space of
a double quantum dot with spin and one orbital per site. Each dot supports the local
states $\ket{0}$, $\ket{\uparrow}$, $\ket{\downarrow}$, and
$\ket{\uparrow\downarrow}$, a description also used in Hubbard-model
studies of correlations in quantum-dot arrays
\cite{Abaach2021,Abaach2022PRA}. The four fermionic modes are ordered as
$(1\uparrow,1\downarrow,2\uparrow,2\downarrow)$ throughout, so the
occupation states and fermionic anticommutation relations fix the phases of all hopping matrix elements:
\begin{align*}
 \ket{\bm n} &= \ket{n_{1\uparrow},n_{1\downarrow},
 n_{2\uparrow},n_{2\downarrow}}, \nonumber\\
 \{c_{i\sigma},c_{j\sigma'}^\dagger\} &= \delta_{ij}\delta_{\sigma\sigma'}, \nonumber\\
 \{c_{i\sigma},c_{j\sigma'}\} &=0,\qquad n_{i\sigma}\in\{0,1\}.
\end{align*}
Charge preparation and charging dynamics can be distinguished by resolving the local
occupations according to total electron number.
Let $\Ne=\sum_{i\sigma}c_{i\sigma}^\dagger c_{i\sigma}$
denote the total electron number and $P_N$ its projector onto the
$N$-electron subspace. 
The projectors satisfy $P_NP_M=\delta_{NM}P_N$ and
$\sum_NP_N=\mathbb I$. Accordingly, the full Hilbert space separates
into five charge sectors,
\begin{equation}
 \mathcal H=\bigoplus_{N=0}^4\mathcal H_N,\qquad
 \mathcal H_N=P_N\mathcal H,\qquad
 d_N=\binom{4}{N},
\label{eq:charge_decomposition}
\end{equation}
with $(d_0,d_1,d_2,d_3,d_4)=(1,4,6,4,1)$. This decomposition
identifies the possible occupations without assuming that a populated
sector is dynamically addressable by a given electrical perturbation.

\subsection{Extended-Hubbard Hamiltonian and electrochemical reference}
\label{subsec:HB}

For semiconductor DQDs, the extended Hubbard model describes charge addition together with
tunneling and hybridization. In particular, the charge-stability diagrams studied by
Yang, Wang, and Das Sarma depend on both on-site and intersite
interactions as well as hopping \cite{Yang2011,Wang2011}, while the
semiconductor-dot array of Hensgens \emph{et al.} demonstrates the
experimental tunability of Fermi--Hubbard parameters
\cite{Hensgens2017}. We retain the minimal form for two equivalent dots and tunneling that preserves spin,
\begin{equation}
 H_B=-t_0\Tb+UD+V\CD+\epsilon\ZD,
\label{eq:HB}
\end{equation}
where the four operators are defined by
\begin{align}
 \Tb&=\sum_{\sigma=\uparrow,\downarrow}
 \left(c_{1\sigma}^\dagger c_{2\sigma}
 +c_{2\sigma}^\dagger c_{1\sigma}\right),\nonumber\\
 D&=\sum_{i=1}^2 n_{i\uparrow}n_{i\downarrow},\nonumber\\
 \CD&=n_1n_2,\qquad n_i=\sum_\sigma n_{i\sigma},\nonumber\\
 \ZD&=\frac{n_1-n_2}{2}.
\label{eq:HBoperators}
\end{align}
Here $t_0>0$ is the static tunnel amplitude, $U$ and $V$ are the
on-site and interdot Coulomb energies, and $\epsilon=\epsilon_1-\epsilon_2$ is the differential
orbital detuning. The common orbital
energy has been chosen as zero, yielding the convention
$\epsilon_1=+\epsilon/2$ and $\epsilon_2=-\epsilon/2$. The model
neglects higher orbitals, direct exchange, pair hopping, spin--orbit
terms, and phonon-assisted processes. Such a truncation provides a
controlled theoretical starting point near the low-energy charge
configurations, but should not be extrapolated indiscriminately to
large detuning: measurements compared with configuration-interaction
calculations reveal important corrections to simple Hubbard exchange
models \cite{Willmes2026}.

The electrostatic terms have a particularly useful exact
reorganization. Since $n_i^2=n_i+2n_{i\uparrow}n_{i\downarrow}$,
the number operator obeys $\Ne(\Ne-1)=2D+2\CD$. Therefore
\begin{equation}
\begin{aligned}
H_B={}&-t_0\Tb+\Ueff D+
 \frac{V}{2}\Ne(\Ne-1)\\
&+\epsilon\ZD,\qquad \Ueff=U-V.
\end{aligned}
\label{eq:HB_reorganized}
\end{equation}
Within a given $N$ sector, the last interaction term contributes the
constant $VN(N-1)/2$, and $\Ueff$ controls the energy cost of double
occupation relative to two singly occupied sites. The term proportional to $V$ must be retained
when comparing different charge sectors, because it contributes to the addition energies and
hence to their thermal populations. By contrast, a uniform one-particle shift
$\epsilon_c\Ne$ leaves every within-sector transition invariant and
amounts to the replacement $\mu\mapsto\mu-\epsilon_c$ in the grand-
canonical ensemble used for preparation.

All contributions to $H_B$ commute with $\Ne$, and the associated
charge symmetry is implemented by $V_\varphi=e^{-i\varphi\Ne}$.
Consequently, $[H_B,\Ne]=0$ and $H_B=\bigoplus_{N=0}^4H_N$,
where $H_N=P_NH_BP_N$.

Separating operations that conserve particle number from unrestricted transformations is
consistent with the treatment of superselection rules in \cite{Bartlett2007,Vaccaro2008}.
The Hamiltonian is invariant under global spin rotations and thus conserves total spin.
Throughout, $U(1)$ refers specifically to the conservation of particle number in our
single-copy ergotropy benchmark; we do not identify this benchmark with the
work extractable using a prescribed collection of electrical gates.

For the spin-independent terms retained in Eq.~\eqref{eq:HB},
the total-spin operators satisfy
$[H_B,\mathbf S^2]=[\Tb,\mathbf S^2]=[\ZD,\mathbf S^2]=0$.
The same identities hold for each driven and stochastic Hamiltonian.
This additional conservation law will be used to refine the extraction
benchmark, while the $U(1)$ definition continues to quantify the
charge-conserving optimum.

At symmetric detuning the Hamiltonian acquires a further discrete
symmetry under interchange of the dots. Let $\mathcal P$ exchange the
fermion operators, so that
$\mathcal P c_{1\sigma}\mathcal P^\dagger=c_{2\sigma}$ and
$\mathcal P c_{2\sigma}\mathcal P^\dagger=c_{1\sigma}$. The two
physically distinct electrical perturbations transform according to
\begin{equation}
 \begin{aligned}
 \mathcal P H_B(\epsilon)\mathcal P^\dagger&=H_B(-\epsilon),\\
 \mathcal P\Tb\mathcal P^\dagger&=\Tb,\\
 \mathcal P\ZD\mathcal P^\dagger&=-\ZD.
 \end{aligned}
\label{eq:parity_selection}
\end{equation}
Thus the barrier operator preserves parity whereas differential
detuning changes it. This distinction parallels the barrier-versus-tilt
control considered for spin-qubit exchange gates
\cite{Reed2016,Martins2016,Shim2018}, although the selection rules
used below follow directly from Eq.~\eqref{eq:parity_selection} and
are not inferred from those experimental protocols.

\begin{figure}[!ht]
\centering
\includegraphics[width=0.98\columnwidth]{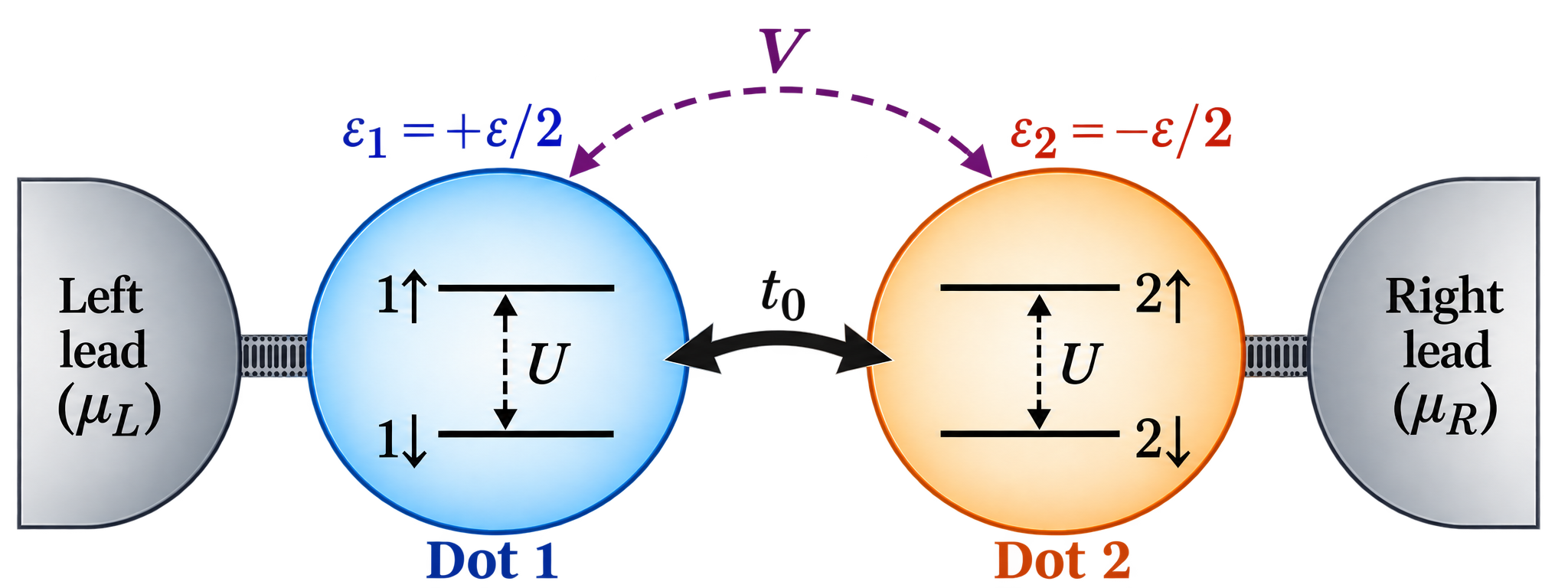}
\caption{\textbf{Two-site spinful extended-Hubbard double quantum dot and preparation protocol.}
The two dots are coupled by the static hopping amplitude $t_0$, with on-site
repulsion $U$, interdot interaction $V$, and orbital detuning
$\epsilon_1=+\epsilon/2$ and $\epsilon_2=-\epsilon/2$.
During grand-canonical loading, the reservoirs have common chemical potential
$\mu$ and temperature $T$; they are disconnected before the isolated
barrier-modulated charging stroke, which conserves the total electron number.}
\label{fig:dqd_schematic}
\end{figure}

Figure~\ref{fig:dqd_schematic} shows the two dots and their preparation through electronic
reservoirs. The two reservoirs are at the same temperature
and chemical potential during equilibration; all reservoir couplings are
switched off during the subsequent number-conserving charging stroke.

\subsection{Exact spectrum and spin structure of the charge sectors}
\label{subsec:sector_structure}

The sector decomposition makes it possible to determine the effects of hopping and interactions
without further truncating the spectrum. The empty and completely filled configurations are
unique:
$\mathcal H_0=\mathrm{span}\{\ket{0,0}\}$ and
$\mathcal H_4=\mathrm{span}\{\ket{\uparrow\downarrow,
\uparrow\downarrow}\}$. Their energies are $E_0=0$ and
$E_4=2U+4V$, so neither sector supports an internal
number-conserving transition.

A single electron, or equivalently a single hole relative to full
occupation, occupies a two-site orbital doublet with a twofold spin
degeneracy. In the occupation ordering already fixed, the $N=1$
doublet is spanned by $\ket{\sigma,0}$ and $\ket{0,\sigma}$.
The corresponding $N=3$ orbitals may be chosen as
$\ket{\uparrow\downarrow,\sigma}$ and
$-\ket{\sigma,\uparrow\downarrow}$; the relative minus sign
fixes the same hopping-matrix convention for both charge sectors.
The resulting Hamiltonians and energy splittings are
\begin{align}
 H_{1,\sigma}&=\begin{pmatrix}
 \epsilon/2&-t_0\\-t_0&-\epsilon/2
 \end{pmatrix},\nonumber\\
 H_{3,\sigma}&=(U+2V)\mathbb I_2+H_{1,\sigma},\nonumber\\
 E_{1,\pm}&=\pm\frac{\Omega}{2},\qquad
 E_{3,\pm}=U+2V\pm\frac{\Omega}{2},\nonumber\\
 \Omega&=\sqrt{\epsilon^2+4t_0^2}.
\label{eq:H13}
\end{align}
Each listed energy is twofold spin degenerate. At $\epsilon=0$,
$H_{1,\sigma}$ and $H_{3,\sigma}$ are affine functions of the
restriction of $\Tb$, a property that suppresses energy absorption
from a barrier-only pulse in these sectors.

Half filling is different because tunneling mixes configurations with
one electron on each site and configurations with a doublon. Spin
rotation symmetry separates the six-dimensional $N=2$ manifold into
three triplets and three singlets. The triplets
$\ket{T_+}=\ket{\uparrow,\uparrow}$,
$\ket{T_-}=\ket{\downarrow,\downarrow}$, and
$\ket{T_0}=(\ket{\uparrow,\downarrow}+
\ket{\downarrow,\uparrow})/\sqrt2$ satisfy
$\Tb\ket{T_m}=0$ and carry the common energy $E_T=V$.
The singlets contain the interdot spin singlet
$\ket S=(\ket{\uparrow,\downarrow}-
\ket{\downarrow,\uparrow})/\sqrt2$ and the localized
doublons $\ket{D_L}=\ket{\uparrow\downarrow,0}$,
$\ket{D_R}=\ket{0,\uparrow\downarrow}$. Their Hamiltonian
has the explicit structure
\begin{equation}
 H_{2,S}=
 \begin{pmatrix}
 U+\epsilon&-\sqrt2t_0&0\\
 -\sqrt2t_0&V&-\sqrt2t_0\\
 0&-\sqrt2t_0&U-\epsilon
 \end{pmatrix}_{\{\ket{D_L},\ket S,\ket{D_R}\}}.
\label{eq:H2singlet}
\end{equation}
The vanishing matrix elements between singlets and triplets are
consequences of spin conservation; the relative tunneling amplitudes
follow from fermionic statistics and the normalization of $\ket S$.

Exchanging the two dots reveals the charging channel selected by the barrier drive.
The states $\ket S$ and
$\ket{D_+}=(\ket{D_L}+\ket{D_R})/\sqrt2$ are even
under $\mathcal P$, whereas
$\ket{D_-}=(\ket{D_L}-\ket{D_R})/\sqrt2$ is odd.
At $\epsilon=0$, the singlet Hamiltonian therefore separates into a
hybridized even doublet and an isolated odd doublon,
\begin{equation}
 H_{2,S}(0)=V\mathbb I_3+
 \begin{pmatrix}
 0&-2t_0&0\\
 -2t_0&\Ueff&0\\
 0&0&\Ueff
 \end{pmatrix}_{\{\ket S,\ket{D_+},\ket{D_-}\}}.
\label{eq:H2parity}
\end{equation}
The even subspace has the eigenvalues
$E_\pm^{(2)}=(U+V\pm\Delta_{\mathrm{eff}})/2$, where
$\Delta_{\mathrm{eff}}=\sqrt{\Ueff^2+16t_0^2}$; the odd doublon
has $E_{D_-}=U$. An exact relation between the singlet--triplet
thermal gap and the upper singlet--odd doublon splitting is
\begin{equation}
 E_T-E_-^{(2)}=E_+^{(2)}-E_{D_-}
 =\frac{\Delta_{\mathrm{eff}}-\Ueff}{2}>0
 \qquad(t_0\ne0).
\label{eq:singlet_gap}
\end{equation}
The smaller gap in Eq.~\eqref{eq:singlet_gap} controls the thermal occupation of the triplets
and the separation between the upper singlet and the odd doublon. The barrier drive addresses
the two even singlets, whose energy difference is $\Delta_{\mathrm{eff}}$.

\label{sec:preparation}

\subsection{Operational charge sectors and equilibrium loading}
\label{subsec:twirling}

Conservation of particle number constrains operations without imposing
an additional restriction on the mathematical form of an arbitrary
input density operator. For energy and work measurements involving only
operators that commute with $\Ne$, coherence between different charge
sectors is inaccessible. The corresponding $U(1)$ group average, or
twirling operation, is \cite{Bartlett2007}
\begin{equation*}
 \GU(\rho)=\frac1{2\pi}\int_0^{2\pi}\!d\varphi\,
 V_\varphi\rho V_\varphi^\dagger
 =\sum_{N=0}^{4}P_N\rho P_N.
\end{equation*}
For any extraction unitary $U$ satisfying $[U,\Ne]=0$,
the work expectation depends only on $\GU(\rho)$: both
$H_B$ and $U^\dagger H_BU$ are block diagonal.

We may consequently represent the operational state by
$\rho_{\mathrm{op}}=\bigoplus_Np_N\rho_N$, with
$p_N=\Tr(P_N\rho)$ and
$\rho_N=P_N\rho P_N/p_N$ for $p_N>0$; sectors with $p_N=0$
make no contribution. This notation separates the loading probability
of a charge sector from its normalized conditional state.

A DQD weakly connected to equilibrium electron reservoirs may exchange
particles during preparation; under equilibration with a common
chemical potential $\mu$ and temperature $T$ (with $\beta=(k_BT)^{-1}$), its equilibrium state
is described by the Gibbs construction
\cite{vanDerWiel2002,Wang2011}
\begin{equation}
\begin{aligned}
 \rho_{\beta,\mu}&=\frac{e^{-\beta(H_B-\mu\Ne)}}{\Xi}, \\
 \Xi&=\Tr e^{-\beta(H_B-\mu\Ne)}.
\end{aligned}
\label{eq:grand_theorem}
\end{equation}
This construction assumes equilibrium loading with a common chemical potential, rather than a
biased state with $\mu_L\ne\mu_R$. Once the reservoirs are decoupled, the sector
weights inherited from preparation remain constant during an isolated,
number-conserving charging operation. The commutator $[H_B,\Ne]=0$
then gives the exact decomposition
\begin{equation}
 \begin{gathered}
\rho_{\beta,\mu}=\bigoplus_{N=0}^{4}p_N\rho_{\beta,N},
 \qquad
 \rho_{\beta,N}=\frac{e^{-\beta H_N}}{Z_N},\\
 Z_N=\Tr_{\mathcal H_N}e^{-\beta H_N},\qquad
 p_N=\frac{e^{\beta\mu N}Z_N}{\Xi},\\
 \Xi=\sum_{N=0}^{4}e^{\beta\mu N}Z_N.
 \end{gathered}
\label{eq:grand_blocks}
\end{equation}
A strictly postselected charge preparation instead corresponds to one
conditional thermal state with $p_{N_0}=1$; the multi-sector treatment
is relevant when the loading stage is not charge postselected or when
it is tuned near an addition boundary.
\subsection{Partition functions and charge-addition boundaries}
\label{subsec:partition}
The exact energies of the five sectors determine their statistical
weights without a phenomenological adjustment of occupations. At the
symmetric point, the orbital doublets, three degenerate triplets, odd
doublon, and two even singlets yield the sector partition functions
\begin{align}
 Z_0&=1,\nonumber\\
 Z_1&=4\cosh(\beta t_0),\nonumber\\
 Z_2&=3e^{-\beta V}+e^{-\beta U}
 +2e^{-\beta(U+V)/2}
 \cosh\!\left(\frac{\beta\Delta_{\mathrm{eff}}}{2}\right),\nonumber\\
 Z_3&=e^{-\beta(U+2V)}Z_1,\nonumber\\
 Z_4&=e^{-\beta(2U+4V)}.
\label{eq:ZN}
\end{align}
A particle--hole relation follows directly from these expressions.
At $\mu_{\mathrm{ph}}=(U+2V)/2$, the grand-canonical weights satisfy
$p_0=p_4$ and $p_1=p_3$, providing exact consistency checks on the
full-Fock-space calculation.
The experimentally familiar charge-addition structure emerges in the
low-temperature limit from the competition among the quantities
$E_{g,N}-\mu N$. In the repulsive regime $\Ueff>0$ the relevant
ground-state energies are
$E_{g,1}=-t_0$,
$E_{g,2}=(U+V-\Delta_{\mathrm{eff}})/2$, and
$E_{g,3}=U+2V-t_0$. Equal grand-canonical energies give the
neighboring two-electron addition boundaries
\begin{align}
 \mu_{12}^{(0)}&=
 t_0+\frac{U+V-\Delta_{\mathrm{eff}}}{2},\nonumber\\
 \mu_{23}^{(0)}&=
 \frac{U+3V+\Delta_{\mathrm{eff}}}{2}-t_0.
\label{eq:mu_boundaries}
\end{align}
Their midpoint equals $\mu_{\mathrm{ph}}$. At finite temperature,
setting neighboring statistical weights equal yields
\begin{equation}
 \mu_{12}^{(T)}=\frac{1}{\beta}\ln\frac{Z_1}{Z_2},\qquad
 \mu_{23}^{(T)}=\frac{1}{\beta}\ln\frac{Z_2}{Z_3}.
\label{eq:mu_thermal}
\end{equation}
These equal-weight conditions are operationally useful because they
allow spectator and active sectors to be appreciably populated
simultaneously without changing the temperature merely to increase
the visibility of higher-charge states.
\subsection{Thermal occupation of inactive singlet and triplet states}
\label{subsec:thermal_spectators}
The probability of occupying a charge sector does not determine how much energy it can absorb.
At fixed $N=2$, a Gibbs preparation
may populate the three spin triplets and the odd doublon in addition to
the two parity-even singlets. Those four states are spectators under the
barrier-only drive at exact symmetry, and their conditional probability is
\begin{equation}
P_{\mathrm{sp}}^{(2)}
=\frac{3e^{-\beta V}+e^{-\beta U}}{Z_2}.
\label{eq:thermal_spectator_population}
\end{equation}
The common spectral gap in Eq.~\eqref{eq:singlet_gap} determines the
thermal accessibility of the triplets and, independently, the proximity
of the upper singlet to the parity-odd doublon. The two effects must be
retained when assessing interaction-dependent charging at finite temperature.

Exchanging the populations of the two even singlets defines an ideal reference that does not
depend on pulse calibration.
For the initial conditional Gibbs state, the population difference is
$(e^{-\beta E_-^{(2)}}-e^{-\beta E_+^{(2)}})/Z_2$ and the resulting
energy increase, equal to $U(1)$ ergotropy for this inversion, is
\begin{equation}
 W_{\mathrm{inv}}^{\max}=p_2\Delta_{\mathrm{eff}}
 \frac{e^{-\beta E_-^{(2)}}-e^{-\beta E_+^{(2)}}}{Z_2}.
\label{eq:thermal_inversion_bound}
\end{equation}
Equation~\eqref{eq:thermal_inversion_bound} describes complete population inversion in the
active doublet. It does not establish that a pulse of finite duration reaches this endpoint. It
also shows why the transition strength $M_{\mathrm{tun}}$ is not
by itself a figure of merit: the thermal preparation can remove population
from the active doublet without altering the barrier matrix element.

\begin{figure*}[t]
\centering
\includegraphics[width=0.98\textwidth]{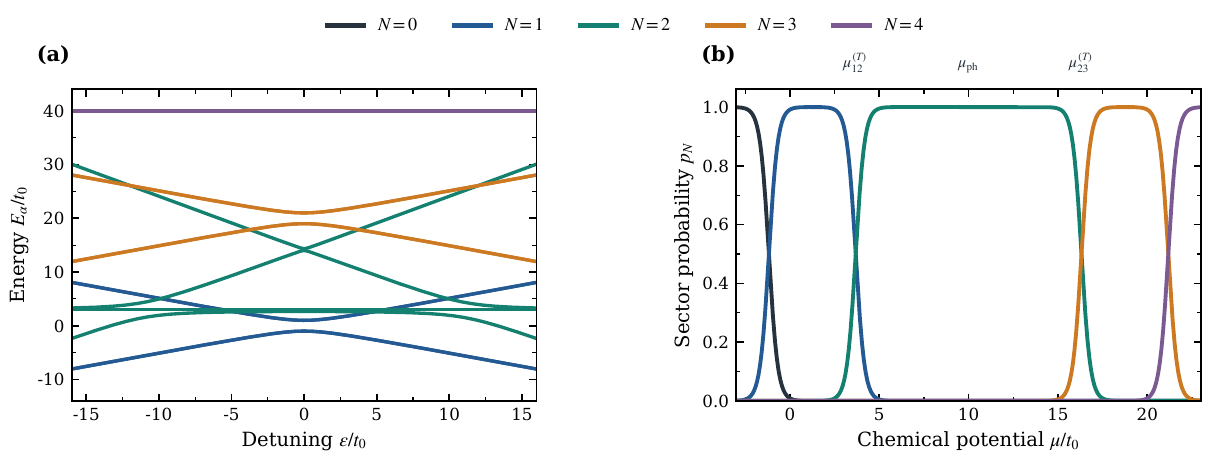}
\caption{\textbf{Many-body spectrum and grand-canonical charge-sector loading.}
(a) Exact energies $E_\alpha/t_0$ versus differential detuning
$\epsilon/t_0$, colored by electron number $N=0,\ldots,4$; the vertical
dashed line indicates $\epsilon=0$.
(b) Equilibrium charge-sector probabilities $p_N(\mu,T)$ at
$k_BT/t_0=0.25$, with equal-weight boundaries
$\mu_{12}^{(T)}$ and $\mu_{23}^{(T)}$ and particle--hole point
$\mu_{\rm ph}=U/2+V$. In both panels $U/t_0=14$ and $V/t_0=3$.
The temperature in (b) makes the loading transitions visible;
the reference charging protocol in Figs.~\ref{fig:ideal_charging} and
\ref{fig:mixing_loss} uses $k_BT/t_0=0.05$.}
\label{fig:charge_preparation}
\end{figure*}

Figure~\ref{fig:charge_preparation}(a) displays the exact many-body
spectrum, while Fig.~\ref{fig:charge_preparation}(b) gives the
charge-sector probabilities at $k_BT/t_0=0.25$, chosen to make the
finite-temperature crossovers visible. The reference preparation used
for the dynamical calculations instead has $k_BT/t_0=0.05$ and is tuned to $\mu=\mu_{12}^{(T)}$, where
$p_1=p_2$ to the plotted precision and the remaining sector weights
are negligible. The thermal population of the spectator states and the
ideal inversion work are specified analytically in
Eqs.~\eqref{eq:thermal_spectator_population} and
\eqref{eq:thermal_inversion_bound}.

\section{Symmetry-constrained work, electrical selection rules, and noise framework}
\label{sec:work}

\subsection{Single-copy ergotropy under particle-number conservation}
\label{subsec:ergotropy_U1}

The ergotropy of a finite quantum system is defined through cyclic
unitary work extraction relative to a fixed Hamiltonian
\cite{Pusz1978,Lenard1978,Allahverdyan2004}. The relevant
optimization depends on the admissible operations: in an isolated
normal DQD the electrical Hamiltonians considered here do not change
$N$, and a charge-changing extraction would require an additional
particle reservoir. The work of a unitary operation $U$ is
\begin{equation}
 W(\rho;U)=\Tr(H_B\rho)
 -\Tr(H_BU\rho U^\dagger).
\label{eq:work}
\end{equation}
Maximizing this expression over all unitary operations that conserve particle number defines
\begin{equation}
 \begin{aligned}
 \mathfrak U_{U(1)}&=\{U\in U(16):[U,\Ne]=0\},\\
 \Wu(\rho)&=\max_{U\in\mathfrak U_{U(1)}}W(\rho;U).
 \end{aligned}
\label{eq:WU}
\end{equation}
This defines a single-copy optimum, distinct from the stronger
requirement of complete passivity involving arbitrarily many copies.

Every $U\in\mathfrak U_{U(1)}$ acts as $\bigoplus_NU_N$, with
$U_N\in U(d_N)$. Because the work expectation is block diagonal, the optimization
is separable. This is the $U(1)$ specialization of the general
single-copy factorization derived in Proposition S1 of the
Supplemental Material of Ref.~\cite{Mitsuhashi2022}:
\begin{equation}
 \Wu(\rho)=\sum_{N=0}^{4}p_N\WN(\rho_N).
\label{eq:WU_factorization}
\end{equation}
More explicitly, for conditional eigenvalues
$q_{1|N}\geq\cdots\geq q_{d_N|N}$ and Hamiltonian energies
$E_{1|N}\leq\cdots\leq E_{d_N|N}$, the sector passive energy and
ergotropy are
\begin{align}
 E_{\mathrm{pass},N}(\rho_N)
 &=\sum_{j=1}^{d_N}q_{j|N}E_{j|N},\nonumber\\
 \WN(\rho_N)
 &=\Tr(H_N\rho_N)-E_{\mathrm{pass},N}(\rho_N).
\label{eq:passive_sector}
\end{align}
Because the edge sectors are one dimensional,
$\mathcal W_0=\mathcal W_4=0$ for any state. In particular,
Eq.~\eqref{eq:WU_factorization} concerns the preparation statistics
$p_N$ and the work stored conditionally in each charge block, not
merely an average over arbitrary qubit occupations.

\subsection{Relation to unrestricted ergotropy}
\label{subsec:work_hierarchy}

The choice of admissible unitary transformations is essential when
comparing energetic benchmarks. An unrestricted unitary may redistribute
populations across different charge sectors, whereas a unitary commuting
with $\Ne$ cannot. If $r_1\geq\cdots\geq r_{16}$ are the eigenvalues of the
full density matrix and $E_1\leq\cdots\leq E_{16}$ are the static
Hamiltonian energies, the unrestricted ergotropy reads
\begin{equation}
\Wg(\rho)=\Tr(H_B\rho)-\sum_{j=1}^{16}r_jE_j.
\label{eq:Wglob}
\end{equation}
Set inclusion immediately yields $0\leq\Wu(\rho)\leq\Wg(\rho)$.

Consequently, $\Wg-\Wu$ is the energetic advantage that would be
available if the particle-number restriction were removed. Ergotropy optimized under $U(1)$
symmetry is an upper bound on the work obtainable using a prescribed set of gates that conserve
particle number.
We do not optimize such gate sequences here.
Adding a uniform electrochemical term $\lambda\Ne$ to the
static Hamiltonian leaves $\Wu(\rho)$ unchanged,
since every admissible operation preserves the total charge. The analogous
statement does not generally hold for unrestricted ergotropy, which allows
particle-number-changing unitary rearrangements. This invariance provides
an additional exact check on the choice of electrochemical reference.

\subsection{Spin-preserving refinement of the work benchmark}
\label{subsec:spin_work}

The microscopic dynamics conserve the total-spin Casimir
$\mathbf S^2$ as well as $\Ne$. This motivates a nested extraction
class containing only the unitaries that respect both conserved quantities,
\begin{equation}
\begin{aligned}
\mathfrak U_{N,S}
 &=\{\mathcal U\in U(16):
 [\mathcal U,\Ne]=[\mathcal U,\mathbf S^2]=0\},\\
\WNS(\rho)
 &=\max_{\mathcal U\in\mathfrak U_{N,S}}
 W(\rho;\mathcal U).
\end{aligned}
\label{eq:spin_work_class}
\end{equation}
Let $P_{N,S}$ be the joint charge--spin projectors, and let
$p_{N,S}=\Tr(P_{N,S}\rho)$ and
$\rho_{N,S}=P_{N,S}\rho P_{N,S}/p_{N,S}$ for
$p_{N,S}>0$. Since the admissible unitaries act independently in
every such subspace, their optimization gives
\begin{equation}
\WNS(\rho)=\sum_{N,S}p_{N,S}
\mathcal W_{N,S}^{\mathrm{block}}(\rho_{N,S}),
\label{eq:spin_factorization}
\end{equation}
where each block ergotropy follows by ordering the eigenvalues of
the conditional state against the increasing energies of its block
Hamiltonian. Inclusion of the admissible unitary sets establishes
\begin{equation}
0\le\WNS(\rho)\le\Wu(\rho)\le\Wg(\rho).
\label{eq:spin_work_hierarchy}
\end{equation}

The extra spin restriction is nontrivial at half filling. In $N=2$,
the three singlets ($S=0$) and three triplets ($S=1$) form distinct
spin sectors. When only particle number is conserved, the passive rearrangement may exchange
populations across the six states of $N=2$. Requiring conservation of total spin prohibits
exchanges between the singlet and triplet subspaces.
In $N=1,3$, all states have $S=1/2$, and in $N=0,4$ the sectors are
one dimensional. Thus only $N=2$ distinguishes these two particular
benchmarks for the present device.

The class in Eq.~\eqref{eq:spin_work_class} conserves $S^2$ but, in
general, permits more unitaries than full $SU(2)$ spin symmetry.
For the states actually considered in this work, however, the
spin-independent Hamiltonian, thermal preparation, barrier control,
and detuning fluctuations preserve invariance under all global spin
rotations. In that restricted family, the $(N,S)$ optimum is attained
by an $SU(2)$-invariant unitary, and the two symmetry benchmarks
coincide; a proof is given in Appendix~\ref{app:spin}
\cite{Bartlett2007,FultonHarris1991}. This equivalence does not imply
controllability using only the available electrical gates, so the
reported work still provides an upper extraction benchmark.

\label{sec:selection}

\subsection{Exact singlet--triplet work deficit}
\label{subsec:exact_spin_work_deficit}

We apply the principle of passive spectral rearrangement to the six states at half filling.
Consider a conditional state $\rho_2$ with $\Tr\rho_2=1$ that is invariant under global
rotations of the spin. Its triplet block is
$q\mathbb I_3$, where $q$ is the population of each triplet state.
Define the singlet block $\rho_s=P_{2,0}\rho_2P_{2,0}$, regarded as an
operator on the three-dimensional singlet subspace. Its ordered
eigenvalues $s_1\ge s_2\ge s_3\ge0$ satisfy
\begin{equation}
\Tr\rho_s=1-3q,\qquad s_1+s_2+s_3+3q=1.
\label{eq:singlet_subnormalization}
\end{equation}
When $1-3q>0$, the normalized conditional singlet state defined in
Sec.~\ref{subsec:spin_work} is
$\rho_{2,S=0}=\rho_s/(1-3q)$. The $s_i$ entering the work-deficit
criterion are the eigenvalues of the \emph{subnormalized} singlet block.

With $U>V$ and $t_0>0$, the ordered six-level energy list is
$(E_-^{(2)},V,V,V,U,E_+^{(2)})$, where
$E_-^{(2)}<V<U<E_+^{(2)}$. Define the spin-work deficit
$\mathcal G_{\rm ST}(\rho)=\Wu(\rho)-\WNS(\rho)$. Direct passive
spectral rearrangement \cite{Bhatia1997} gives
\begin{equation}
\begin{split}
\mathcal G_{\rm ST}(\rho)=p_2\big[&
(V-E_-^{(2)})(q-s_1)_+\\
&+(U-V)(s_2-q)_+\\
&+(E_+^{(2)}-V)(s_3-q)_+\big],
\end{split}
\label{eq:exact_ST_gap}
\end{equation}
where $(x)_+=\max\{x,0\}$. Appendix~\ref{app:ST_closed_form} gives the proof by ordering the
six eigenvalues. For $p_2>0$, the
deficit vanishes if and only if $s_2\le q\le s_1$. In the branch
$q<s_1$, its appearance requires $s_2>q$. Equality of the two work
benchmarks does not make the admissible unitary classes identical:
the spin restriction remains exact but is nonbinding for such a state.

For example, if $s_1\ge s_2>q\ge s_3$, passive rearrangement that preserves spin assigns $s_2$
to the singlet energy $U$. With only
particle number conserved, $s_2$ can instead occupy a triplet
energy $V$, while a triplet weight $q$ is assigned to $U$. Therefore
$\mathcal G_{\rm ST}=p_2(U-V)(s_2-q)$. The difference compares two classes of unitary
extraction; no physical gate that transfers population between the spin sectors is assumed. The
identity follows from established results on passive spectral rearrangement applied to the
spectrum of this model.

\subsection{Barrier control in the isolated device}
\label{subsec:barrier_drive}

Electrically tunable interdot barriers provide a natural control
parameter for charge hybridization. Symmetric barrier operation has
been established in semiconductor spin qubits, where it can reduce
sensitivity to detuning-like fluctuations
\cite{Reed2016,Martins2016,Shim2018}. To examine the selection rules associated with charge conservation, we model a controlled modulation
of the hopping amplitude by
\begin{equation}
 \begin{aligned}
 t(t)&=t_0+\delta t(t),\\
 \delta t(t)&=A_tg(t)\cos(\omega_dt+\phi),\\
 H(t)&=H_B-\delta t(t)\Tb.
 \end{aligned}
\label{eq:tmod}
\end{equation}
Here $g(t)$ specifies the pulse envelope. Since $\Tb$ preserves
particle number, every fixed-$N$ block evolves unitarily, and its
probability $p_N$ remains constant. We evaluate stored energy relative to the static
Hamiltonian $H_B$, rather than the instantaneous Hamiltonian during charging.

An important thermodynamic consequence of isolated charging is that
the eigenvalues of every conditional state $\rho_N(t)$ are constant.
Its sector passive energy is therefore invariant under the evolution,
and Eq.~\eqref{eq:WU_factorization} immediately gives
\begin{equation}
 \Wu[\rho(t)]-\Wu(\rho_0)
 =\Tr\!\left[H_B(\rho(t)-\rho_0)\right]
 \equiv\Est(t).
\label{eq:W_energy_identity}
\end{equation}
If the initial equilibrium state is passive with respect to
number-conserving operations, $\Wu(\rho_0)=0$, and the stored energy
coincides exactly with the symmetry-constrained ergotropy. It does not imply that the optimum
can be reached by a prescribed discharge waveform.

\subsection{Interaction-induced selection in the half-filled sector}
\label{subsec:selection_rules}

The commutator between the static Hamiltonian and the driving operator determines which sectors
can absorb energy. At $\epsilon=0$, the one- and
three-electron Hamiltonians in Eq.~\eqref{eq:H13} are affine
functions of their respective $\Tb$ blocks. Thus
$[H_N(0),\Tb]=0$ for $N=1,3$, while $N=0,4$ do not support
internal transitions. Although those sectors can carry nonzero loading
weights, a barrier pulse cannot change their static energy under
this symmetric, noiseless protocol.

At half filling, however, singlet tunneling incurs an interaction-dependent energy cost. Its
parity-even two-level subsystem, identified in
Eq.~\eqref{eq:H2parity}, does not generally commute with the barrier
operator. The two distinct symmetry channels are exposed directly by
\begin{equation}
 \Tb^{(2,S)}=
 \begin{pmatrix}0&2&0\\2&0&0\\0&0&0\end{pmatrix},
 \qquad
 \ZD^{(2,S)}=
 \begin{pmatrix}0&0&0\\0&0&1\\0&1&0\end{pmatrix},
\label{eq:drive_noise_operators}
\end{equation}
written with the parity-adapted ordering already defined above.
The barrier operator does not couple to $\ket{D_-}$, while detuning connects states of opposite
parity. For the even subspace the static- and drive-operator
commutator is
\begin{equation}
 [H_{2,S}(0),\Tb^{(2,S)}]
 =2\Ueff\begin{pmatrix}
 0&-1&0\\1&0&0\\0&0&0
 \end{pmatrix}.
\label{eq:commutator_N2}
\end{equation}
At $\Ueff=0$, the commutator vanishes. In fact, the exact
reorganization in Eq.~\eqref{eq:HB_reorganized} shows the stronger
statement $[H_B,\Tb]=0$ on the full Fock space when $U=V$ and
$\epsilon=0$. The inability of a barrier-only modulation to change
the static battery energy at this point is therefore not an artifact
of a two-state truncation.

For $\Ueff\neq0$, the even singlet eigenstates
$\ket{E_\pm^{(2)}}$ are coupled by the barrier operator. The squared
transition matrix element follows by diagonalizing the even block,
\begin{equation}
 M_{\mathrm{tun}}
 \equiv\left|\bra{E_+^{(2)}}\Tb\ket{E_-^{(2)}}\right|^2
 =\frac{4\Ueff^2}{\Ueff^2+16t_0^2}.
\label{eq:Mtun}
\end{equation}
The transition strength vanishes quadratically as $U$ approaches $V$ and increases with
$|U-V|$. The matrix element determines the coupling strength. The stored energy and work also
depend on the initial singlet populations and on the pulse duration. For an initially thermal
state and ideal barrier-only
driving at symmetric detuning, the triplets and odd doublon remain
spectators, so $\Wu[\rho(t)]=p_2\mathcal W_2[\rho_2(t)]$.
This equality is protocol specific and does not assert that the
$N=1,3$ sectors remain inert once detuning noise or detuning control
is present.

\subsection{Parity-resolved detuning response}
\label{sec:parity_noise}

Barrier modulation and differential detuning preserve the electron number but select different transitions.
The parity-odd operator $\ZD$ couples the odd doublon only to the
even doublon component of the singlets. The two squared matrix
elements are
\begin{equation}
 M_{\epsilon,\pm}
 =\left|\bra{D_-}\ZD\ket{E_\pm^{(2)}}\right|^2
 =\frac12\left(1\pm\frac{\Ueff}{\Delta_{\mathrm{eff}}}\right).
\label{eq:Mnoise}
\end{equation}
They satisfy $M_{\epsilon,+}+M_{\epsilon,-}=1$. Although the
symmetric point is stationary under a reversal of detuning,
its transition frequencies generally retain a nonzero
second-order sensitivity. For nondegenerate parity eigenstates,
$\langle E_n|\ZD|E_n\rangle=0$, and the
Hellmann--Feynman theorem gives
$\left.\partial_\epsilon E_n\right|_{\epsilon=0}=0$.

Standard second-order perturbation theory then relates the
sweet-spot curvature to allowed matrix elements and the associated
energy denominators,
\begin{equation}
 \left.\frac{\partial^2 E_n}{\partial\epsilon^2}\right|_0
 =2\sum_{m\ne n}
 \frac{|\bra{E_m}\ZD\ket{E_n}|^2}{E_n-E_m}.
\label{eq:second_order_susceptibility}
\end{equation}
For the two even singlets, the only parity-allowed state appearing
in the sum is $\ket{D_-}$. Therefore their curvatures obey
\begin{equation}
 \left.\frac{\partial^2 E_\pm^{(2)}}
 {\partial\epsilon^2}\right|_0
 =\frac{2M_{\epsilon,\pm}}{E_\pm^{(2)}-U}.
\label{eq:singlet_curvature}
\end{equation}
The relevant frequency for the ideal charging pulse is determined by
the \emph{difference} of the two even-singlet energies, rather than by
the curvature of either state alone. Defining
$\Delta_{+-}(\epsilon)=E_+^{(2)}(\epsilon)-E_-^{(2)}(\epsilon)$, direct
subtraction of Eq.~\eqref{eq:singlet_curvature} yields
\begin{equation}
\begin{aligned}
\kappa_{+-}&\equiv
\left.\frac{\partial^2\Delta_{+-}(\epsilon)}
{\partial\epsilon^2}\right|_{\epsilon=0}\\
&=\frac{\Delta_{\mathrm{eff}}^2+\Ueff^2}
{4t_0^2\Delta_{\mathrm{eff}}}>0.
\end{aligned}
\label{eq:charging_gap_curvature}
\end{equation}
The singlet gap is insensitive to detuning at first order at the symmetric point, but its
leading response is quadratic. For a
zero-mean Gaussian quasistatic fluctuation of variance
$\sigma_\epsilon^2$, the weak-noise expansion gives
\begin{equation}
\mathbb E[\Delta_{+-}(\delta\epsilon)]
-\Delta_{\mathrm{eff}}
=\frac{\kappa_{+-}}{2}\sigma_\epsilon^2
+O(\sigma_\epsilon^4).
\label{eq:average_gap_shift}
\end{equation}
This expression describes the shift in the mean transition gap; it does not determine the loss
of ergotropy caused by ensemble mixing. Its perturbative use requires the
fluctuation amplitude to be small compared with the relevant
parity-opposite level separations. This microscopic charging-gap
response complements the symmetric-operation and barrier-versus-tilt
analyses of semiconductor spin qubits
\cite{Reed2016,Martins2016,Shim2018}, without identifying the present
singlet--singlet charging transition with a singlet--triplet exchange
qubit.
The proximity of the upper singlet and odd doublon is important:
when $\Ueff/t_0\gg1$, their separation satisfies
$E_+^{(2)}-U\simeq4t_0^2/\Ueff$ while
$M_{\epsilon,+}\to1$. Hence increased interaction enhances the
barrier transition but can also make its upper level more
susceptible to low-frequency detuning fluctuations. Matrix elements and energy gaps alone do
not determine the response to a charging pulse of finite duration. In the quasistatic protocol
below, each offset
modifies the coherent propagator, and the mixing loss results from
averaging the final states. It is not an irreversible transition rate.

For the reference interaction $U/t_0=14$ and $V/t_0=3$,
the upper even singlet and odd doublon have the exact separation
$(\sqrt{137}-11)t_0/2\simeq0.35235t_0$. The associated small
energy denominator enhances the curvature in
Eq.~\eqref{eq:singlet_curvature}. This local spectral sensitivity
alone does not prove that the stored or extractable work is
nonmonotonic in the interaction: locating a performance optimum requires examining the charged
state over a range of interaction strengths after averaging over noise.

At $U/t_0=14$ and $V/t_0=3$ one has
$\Delta_{\mathrm{eff}}/t_0=\sqrt{137}$,
$M_{\mathrm{tun}}\simeq3.533$, $M_{\epsilon,+}\simeq0.970$,
and $M_{\epsilon,-}\simeq0.030$.
\begin{figure*}[t]
\centering
\includegraphics[width=0.98\textwidth]{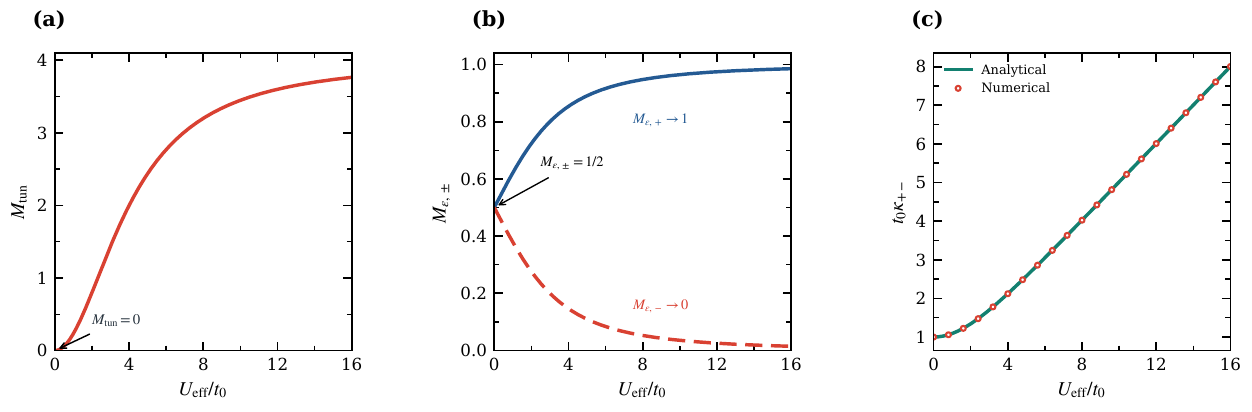}
\caption{\textbf{Interaction-enabled barrier coupling and quadratic detuning response.}
(a) Squared barrier matrix element $M_{\rm tun}$ between the two parity-even
singlets. (b) Squared detuning matrix elements $M_{\epsilon,+}$ and
$M_{\epsilon,-}$ connecting the even singlets to the odd doublon.
(c) Dimensionless curvature $t_0\kappa_{+-}$ of the even-singlet energy
gap: analytical expression (solid curve) and finite-difference
results from direct diagonalization (markers).
All panels use $\epsilon=0$ and scan
$U_{\rm eff}/t_0=(U-V)/t_0$. The panels characterize the spectrum and its response to detuning, rather than the ergotropy after averaging over noise.}
\label{fig:interaction_selectivity}
\end{figure*}

Figure~\ref{fig:interaction_selectivity} compares the barrier coupling, parity-allowed detuning
matrix elements,
and the curvature of the driven even-singlet gap. The agreement between
Eq.~\eqref{eq:charging_gap_curvature} and direct finite-difference
diagonalization provides an independent check on the interaction-dependent
quadratic susceptibility.

\subsection{Quasistatic detuning fluctuations and ensemble averaging}
\label{sec:noise}

\subsubsection{Stochastic Hamiltonian and noise convention}
\label{subsec:noise_model}

Electrical fluctuations in semiconductor quantum dots commonly exhibit
low-frequency colored spectra, with device- and bandwidth-dependent
magnitudes and exponents \cite{Dial2013,Paladino2014,Connors2019,
Connors2022,Burkard2023}. We model differential detuning as a real stationary classical
stochastic process without imposing Markovian dynamics. Our
one-sided spectral convention is fixed by
\begin{equation}
 \begin{gathered}
 S_\epsilon(f)=
 \begin{cases}
 A_\epsilon f^{-\alpha},&f_l\le f\le f_h,\\
 0,&\text{otherwise},
 \end{cases}\\
 \big\langle\delta\epsilon(t)\delta\epsilon(t+\tau)\big\rangle
 =\int_0^\infty S_\epsilon(f)\cos(2\pi f\tau)\,df.
 \end{gathered}
\label{eq:PSD}
\end{equation}
Here $f_l$ and $f_h$ are explicit infrared and ultraviolet
cutoffs. In the idealized $1/f$ case $\alpha=1$, the integrated
detuning variance is
$\sigma_\epsilon^2=A_\epsilon\ln(f_h/f_l)$.
A quoted rms amplitude must therefore be accompanied by its spectral bandwidth. With local
energy fluctuations $\delta\mu_i$,
the differential noise spectrum satisfies
$S_\epsilon(f)=S_{11}(f)+S_{22}(f)-2\operatorname{Re}S_{12}(f)$,
where $S_{ij}$ denotes the cross-spectral density, with the same
normalization convention as the one-sided auto-spectra. The differential noise amplitude
depends on correlations between the dots and cannot be inferred from measurements at one dot
alone.

Each realization of the random process gives a unitary evolution
generated by the microscopic Hamiltonian
\begin{equation}
 H_r(t)=H_B-\delta t(t)\Tb
 +\delta\epsilon_r(t)\ZD,
 \qquad [H_r(t),\Ne]=0.
\label{eq:Hnoise}
\end{equation}
The static battery energy is not conserved by the individual
random Hamiltonians. Its rate of change contains contributions
from both the imposed control and the stochastic detuning,
\begin{equation*}
 \frac{d}{dt}\langle H_B\rangle_r
 =\frac{i}{\hbar}\left\langle
 -\delta t(t)[\Tb,H_B]
 +\delta\epsilon_r(t)[\ZD,H_B]
 \right\rangle_r.
\end{equation*}
Hence detuning noise can modify the mean stored energy as well
as the coherence of the unresolved state. Moreover, $\ZD$ does not
commute with the one- and three-electron hopping Hamiltonians at
$\epsilon=0$: their inactivity under a \emph{barrier-only} pulse
cannot be carried over to noisy evolution.

When the noise realization is not recorded, the operational state is
the ensemble average
\begin{equation}
 \bar\rho(t)=\mathbb E_r\!\left[
 U_r(t)\rho_0U_r^\dagger(t)\right]
 \equiv\Phi_t(\rho_0).
\label{eq:ensemble}
\end{equation}
The channel $\Phi_t$ is an average of unitary maps and is therefore trace preserving and unital.
Each trajectory preserves particle number, so the preparation
weights $p_N$ are unchanged after averaging, although
$\bar\rho_N(t)$ need not be unitarily related to its initial
conditional state.

\paragraph{Parity symmetry restored by the unresolved ensemble.}
Under exchange of the two dots, detuning changes sign, whereas the static Hamiltonian at zero
detuning, the barrier drive, and the initial Gibbs state remain invariant. For a static offset
$\eta=\delta\epsilon/t_0$, let $\mathcal P$ denote dot exchange.
The exact covariance relation
$\mathcal P H_{\eta}(t)\mathcal P^\dagger=H_{-\eta}(t)$ implies
$U_{-\eta}(t)=\mathcal P U_{\eta}(t)\mathcal P^\dagger$.
When the distribution is even, $p(\eta)=p(-\eta)$, the averaged
state consequently satisfies
\begin{equation}
 \mathcal P\bar\rho(t)\mathcal P^\dagger=\bar\rho(t),
 \qquad [\bar\rho(t),\mathcal P]=0.
 \label{eq:averaged_parity}
\end{equation}
A trajectory with a nonzero detuning offset need not preserve parity, but the average over
opposite offsets does. In the parity-adapted $N=2$ singlet basis
$\{|S\rangle,|D_+\rangle,|D_-\rangle\}$, with
$\mathcal P_S=\operatorname{diag}(1,1,-1)$, it gives
\begin{equation}
 \bar\rho_s(t)=
 \begin{pmatrix}
 B_{\rm even}(t)&0\\
 0&\lambda_{\rm odd}(t)
 \end{pmatrix},
 \label{eq:parity_averaged_block}
\end{equation}
Here $\bar\rho_s=P_{2,0}\bar\rho_2P_{2,0}$ is the singlet block of
the normalized conditional state $\bar\rho_2$, so that
$\Tr B_{\rm even}+\lambda_{\rm odd}=1-3q$.
The singlet block $\bar\rho_s$ is not separately normalized.
The positive $2\times2$ matrix $B_{\rm even}$ has two
eigenvalues $\lambda_{{\rm even},+}\ge\lambda_{{\rm even},-}$
and the odd weight $\lambda_{\rm odd}$ form the singlet spectrum
$(s_1,s_2,s_3)$ after sorting. When $\lambda_{{\rm even},-}=\lambda_{\rm odd}$, the parity
sector associated with $s_2$ changes. Spin symmetry remains intact, and the noise distribution
is unchanged. We identify this distinct internal spectral
reordering in the temperature--noise map below.

\subsubsection{Mixing-induced passive-energy increase}
\label{subsec:mixing_loss}

The effect of averaging over unknown noise realizations can be established without specifying a
particular noise spectrum. Since $\Wu$ is the maximum of a family
of linear work functionals, it is convex in the state:
\begin{equation}
\begin{aligned}
 \Wu\!\left(\sum_r\lambda_r\rho_r\right)&\le\sum_r\lambda_r\Wu(\rho_r), \\
 \lambda_r&\ge0,\qquad \sum_r\lambda_r=1.
\end{aligned}
\label{eq:convexity}
\end{equation}
For every noise realization, the evolution is unitary within each
charge sector. Its conditional density-matrix spectra and hence
its symmetry-constrained passive energy remain constant. The
trajectory-resolved ergotropy therefore obeys
$\Wu[\rho_r(t)]-\Wu(\rho_0)
=\Tr[H_B(\rho_r(t)-\rho_0)]$.
Combining this invariance with the convexity inequality gives
\begin{equation}
 \Wu[\bar\rho(t)]-\Wu(\rho_0)
 \le\Tr[H_B(\bar\rho(t)-\rho_0)]
 =\Est(t).
\label{eq:noise_inequality}
\end{equation}

The nonnegative difference measures the rise in passive energy when extraction must act on the
averaged state rather than on a known realization. It can equivalently be viewed as
the advantage of tailoring the optimal extraction to a known classical noise
label rather than operating on an unresolved mixture. This is conceptually
related to information-assisted work extraction~\cite{Francica2017}, but
requires neither ancilla measurements nor quantum correlations. We define it by
\begin{equation}
 \begin{aligned}
 \Lmix(t)&=\Est(t)
 -\{\Wu[\bar\rho(t)]-\Wu(\rho_0)\}\\
 &=E_{\mathrm{pass}}^{U(1)}[\bar\rho(t)]
 -E_{\mathrm{pass}}^{U(1)}(\rho_0)\ge0,
 \end{aligned}
\label{eq:Lmix}
\end{equation}
with $E_{\mathrm{pass}}^{U(1)}(\rho)
=\sum_Np_NE_{\mathrm{pass},N}(\rho_N)$.
The block structure resolves this loss into independent charge
contributions,
\begin{equation}
 \Lmix(t)=\sum_Np_N\left[
 E_{\mathrm{pass},N}(\bar\rho_N(t))
 -E_{\mathrm{pass},N}(\rho_N(0))\right].
\label{eq:Lsector}
\end{equation}
For the initial grand-canonical equilibrium state,
$\Wu(\rho_0)=0$, so Eq.~\eqref{eq:Lmix} simplifies to
$\Lmix(t)=\Est(t)-\Wu[\bar\rho(t)]$.
The resulting loss is defined for the full class of operations allowed by the stated symmetry.
Restrictions imposed by a particular electrical gate set may further reduce extractable work.
In particular, random
unitarity and charge conservation do not imply that the noise
is energetically inert; they guarantee only the nonnegativity
of the \emph{mixing-induced} contribution as defined above.

The refinement imposing spin conservation obeys the same mixing inequality.
Each stochastic propagator commutes with both $\Ne$ and
$\mathbf S^2$, hence it is unitary in every $(N,S)$ block.
The resulting passive energy is invariant along an individual
realization, but may increase after the unobserved realization
is averaged out. Consequently,
\begin{equation}
\begin{aligned}
\LNS(t)&=\Est(t)-
\{\WNS[\bar\rho(t)]-\WNS(\rho_0)\}\\
&=\EpNS[\bar\rho(t)]-\EpNS(\rho_0)\ge0.
\end{aligned}
\label{eq:spin_mix_loss}
\end{equation}
The initial grand-canonical Gibbs state is passive in each
joint charge--spin block, so that
$\Wu(\rho_0)=\WNS(\rho_0)=0$. For the preparation used
below, the nesting of the extraction classes therefore yields
\begin{equation}
\LNS(t)-\Lmix(t)=
\Wu[\bar\rho(t)]-\WNS[\bar\rho(t)]\ge0.
\label{eq:spin_loss_difference}
\end{equation}
The inequality compares extraction optima for the same state and does not introduce a second
dissipation mechanism.
The difference admits a more explicit microscopic resolution. The
charge sectors $N=0,4$ are one dimensional, while $N=1,3$ contain
only total spin $S=1/2$. Thus their passive-energy minima are
identical under the two constraints; only the singlet--triplet
structure of the half-filled sector can distinguish them. Since
the initial conditional Gibbs states are passive for both benchmarks,
Eq.~\eqref{eq:spin_loss_difference} reduces exactly to
\begin{equation}
\begin{aligned}
\LNS(t)-\Lmix(t)
=p_2\bigl[&E_{\mathrm{pass},2}^{N,S}(\bar\rho_2(t))\\
&-E_{\mathrm{pass},2}^{U(1)}(\bar\rho_2(t))\bigr]
\geq 0.
\end{aligned}
\label{eq:N2_spin_mixing_gap}
\end{equation}
Here $E_{\mathrm{pass},2}^{N,S}$ is the conditional minimum energy
obtained by passive rearrangement separately within the $S=0$ and
$S=1$ subspaces; $E_{\mathrm{pass},2}^{U(1)}$ instead permits passive
rearrangement across all six states at $N=2$. This identity locates
the contrast between the two panels of Fig.~\ref{fig:mixing_loss}
\emph{entirely} in the half-filled sector. It is the microscopic
realization of the operational distinction between unrestricted
passivity and symmetry-constrained work extraction
\cite{Allahverdyan2004,Vaccaro2008}.

For the initial Gibbs state, the difference between the mixing losses equals the spin-work
deficit derived in Sec.~\ref{subsec:exact_spin_work_deficit}. Averaging changes the singlet
eigenvalues while leaving the triplet population fixed, and can therefore produce a positive
deficit. Each realization still conserves spin:
the difference concerns optimal passive assignment, not an additional
spin-relaxation process.

\section{Results and discussion}
\label{sec:results_discussion}

We now apply the analytic framework to the parameterized extended-Hubbard
battery and its stochastic dynamics. We first identify the charge sectors that absorb energy
under barrier modulation. We then quantify how unresolved detuning fluctuations raise the
passive energy under the two extraction constraints, and determine how thermal populations
govern the boundary at which the additional spin restriction ceases to reduce the extractable
work. In each case, the interpretation follows directly
from the spectral identities derived above. All energies below are referenced
to the fixed, undriven Hamiltonian $H_B$.

\subsection{Sector-selective charging and thermal work capacity}
\label{sec:ideal_thermal}
\subsubsection{Operating parameters and the charging protocol}
\label{sec:parameters}

The numerical results are expressed primarily in units of the static tunnel
amplitude $t_0$.  This choice separates the dimensionless physics from device-
specific lever arms while allowing the parameter window to be compared with
calibrated quantum dots.  Extended-Hubbard descriptions are experimentally
well established in gate-defined arrays: Hensgens \emph{et al.} extracted
$t$, on-site $U_i$, and intersite $V_{ij}$ directly from charge-stability and
transport data \cite{Hensgens2017}. This experimentally calibrated
Hubbard approach complements quantum descriptions of the addition spectrum
\cite{Yang2011,Wang2011}. Recent spinful QD--superconductor
experiments operate in a regime where charging energies of order
$1$--$2~\mathrm{meV}$ are much larger than interdot couplings of order tens of
$\mu\mathrm{eV}$ \cite{tenHaaf2026}.  These measurements support studying a regime with strong
Coulomb interactions, including a finite interdot repulsion $V$.

As a representative normalization, we use the Hubbard energy scales
illustrated in the three-dot study of Hensgens \emph{et al.}
\cite{Hensgens2017}. Their simulated charge-stability diagram in
Extended Data Fig.~6(b) uses a tunnel coupling of
$0.17~\mathrm{meV}$, on-site energies of
$2.39$--$2.92~\mathrm{meV}$, and intersite Coulomb parameters
$V_{12}=0.55$, $V_{23}=0.47$, and $V_{13}=0.27~\mathrm{meV}$.
The last value describes the two outer dots, not nearest neighbors.
We use these values to set an illustrative energy scale, without fitting the two dot model to
that particular device.
We choose $U/t_0=14$ and $V/t_0=3$, giving
$U_{\mathrm{eff}}/t_0=11$. Figure~\ref{fig:interaction_selectivity}
scans $0\le U_{\mathrm{eff}}/t_0\le16$, while the separate auxiliary
interaction--noise calculation uses $U_{\mathrm{eff}}/t_0=5,11,15$.
The thermal
reference $k_BT/t_0=0.05$ satisfies $k_BT\ll t_0$ but still resolves the
finite-temperature crossover between neighboring charge sectors; temperature--noise scans cover approximately
$0.035\le k_BT/t_0\le0.30$.  These values are not
intended as a fit to one specific device.

The use of a single-orbital extended-Hubbard model requires a corresponding
statement of scope.  Microscopic calculations and experiment show that the
simple Hubbard approximation can be accurate near a charge transition or in a
restricted low-energy window, while excited orbitals and configuration-
interaction physics become important when one attempts quantitative modeling
over a broad detuning range \cite{Wang2011,Willmes2026}.  Our calculations
therefore target the lowest charge configurations around the symmetric working
point and the interaction-enabled selection rule.  They do not claim a
quantitative description of far-detuned exchange, valley excitations, detailed
orbital deformation, spin--orbit coupling, phonon relaxation, or device-
specific gate cross-capacitances.

The control scale is treated separately from the static Hubbard parameters.
Symmetric exchange operation has been demonstrated by modulating the interdot
barrier while keeping the double dot close to equal electrochemical potential;
this suppresses first-order sensitivity to detuning-like voltage noise while
retaining strong tunability \cite{Reed2016,Martins2016,Shim2018}.  We use $A_t/t_0=0.10$ as a
theoretical reference in the displayed
dynamics. The chosen amplitude specifies the model drive; it has not been calibrated for the
proposed singlet transition in an experiment.

The preparation chemical potential is chosen according to the physical
question.  The particle--hole point $\mu_{\mathrm{ph}}=U/2+V$ provides the
exact checks $p_0=p_4$ and $p_1=p_3$, whereas the main multi-sector example is
placed near $\mu_{12}^{(T)}$ so that $N=1$ and $N=2$ are both appreciably
occupied.  This mirrors the logic of charge-stability measurements, in which
gate-controlled electrochemical potentials determine the most probable charge
configuration \cite{vanDerWiel2002,Wang2011}.

A direct one-photon transition between the two hybridized even
singlets has frequency
$f_d=\Delta_{\mathrm{eff}}/h=\sqrt{\Ueff^2+16t_0^2}/h$.

Using the representative parameter values $t_0\simeq0.17~\mathrm{meV}$,
$U\simeq2.39~\mathrm{meV}$, and $V\simeq0.55~\mathrm{meV}$ gives
$U_{\mathrm{eff}}\simeq1.84~\mathrm{meV}$, hence an even-singlet
splitting of approximately $1.96~\mathrm{meV}$, or $0.47~\mathrm{THz}$.
The corresponding value from the simulation ratios $U/t_0=14$ and $V/t_0=3$,
with $t_0=0.17~\mathrm{meV}$, is about $0.48~\mathrm{THz}$.
These frequencies are inferred from illustrative Hubbard parameters; it does not imply that the
transition has been driven in an experiment.  We
therefore use the neutral term \emph{ac barrier modulation}.  At fixed $t_0$, the direct
one-photon frequency obeys
$f_d\geq4t_0/h\simeq0.164~\mathrm{THz}$, with equality at the
interaction-free point where the barrier-transition strength vanishes.
Reducing $U_{\mathrm{eff}}$ alone therefore does not place this one-photon
transition in a conventional few-GHz band. A lower carrier frequency would require a different
device scale or an explicit analysis of multiphoton or nonresonant control.

Implementing this hypothetical direct singlet resonance would
require barrier modulation at the quoted subterahertz frequency,
adequate control bandwidth, calibrated tunnel-amplitude response, and
suppression of parasitic detuning. The Hubbard parameter ranges
reported by Hensgens \emph{et al.} motivate the energy scales but do not
constitute a demonstration of this particular driving protocol.

\subsubsection{Ideal charging and thermal blocking}
\label{sec:ideal_dynamics}

We first isolate the noiseless charging problem before introducing the
stochastic detuning term.  The conditional state in each fixed-$N$ block obeys
{
\begin{equation*}
i\hbar\,\dot\rho_N(t)
=
[H_N(t),\rho_N(t)],
\qquad
H_N(t)=H_N-\delta t(t)\Tb^{(N)},
\end{equation*}
}
with $\omega_d=\Delta_{\rm eff}/\hbar$.  We use the scaled time
$\tau=t t_0/\hbar$, a time step $\Delta\tau=0.01$, and
$A_t/t_0=0.10$.  To avoid an abrupt switch-on, the envelope is chosen as
{
\begin{equation}
g(\tau)=
\begin{cases}
\sin^2\!\left(\dfrac{\pi\tau}{2\tau_r}\right),&0\le\tau<\tau_r,\\[2mm]
1,&\tau\ge\tau_r,
\end{cases}
\qquad \tau_r=1.
\label{eq:drive_envelope_numeric}
\end{equation}
}
The preparation is taken at $\mu=\mu_{12}^{(T)}$.  For the reference
parameters $U/t_0=14$, $V/t_0=3$, and $k_BT/t_0=0.05$, the numerical weights
are $p_1=p_2=1/2$ to the displayed precision, while $p_0$, $p_3$, and $p_4$
are negligible.  The prepared ensemble thus contains both the spectator sector $N=1$ and the
active sector $N=2$. The quantum battery
whose spin-work deficit we diagnose is the conditional $N=2$ system;
preparation at this loading boundary reduces the overall signal by the sector weight $p_2$
without changing the criterion within $N=2$. For a postselected $N=2$ initialization,
$p_2=1$ and the conditional deficit is unchanged.

To resolve the active channel inside $N=2$, we use the conditional
populations
{
\begin{align*}
P_\pm^{(2)}(t)&=\langle E_\pm^{(2)}|\rho_2(t)|E_\pm^{(2)}\rangle,\nonumber\\
P_{\rm sp}^{(2)}(t)&=1-P_-^{(2)}(t)-P_+^{(2)}(t).
\end{align*}
}
where $P_{\rm sp}^{(2)}$ collects the odd doublon and triplet
populations.  At the symmetric barrier-only working point these states are not
coupled to the driven even-singlet doublet, so their population remains
constant.

\begin{figure*}[t]
\centering
\includegraphics[width=0.98\textwidth]{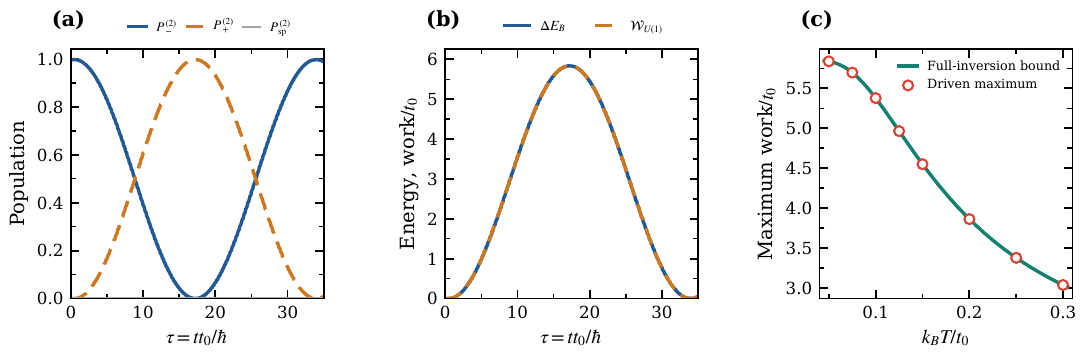}
\caption{\textbf{Charging of the active sector and thermal reduction of extractable work.}
(a) Conditional $N=2$ populations $P_-^{(2)}$ and $P_+^{(2)}$ of the
parity-even singlets and the spectator population $P_{\rm sp}^{(2)}$.
(b) Stored energy $\Delta E_B$ and ergotropy $\mathcal W_{U(1)}$ optimized under particle number conservation for the noiseless barrier pulse; the curves overlap
because the initial Gibbs state is passive and the sectorwise evolution
is unitary. Panels (a) and (b) use $k_BT/t_0=0.05$.
(c) Ideal full-inversion work $W_{\rm inv}^{\max}$ and the maximum
$\mathcal W_{U(1)}$ attained by the specified pulse over
$0\le\tau\le35$, versus preparation temperature.
In all panels $U/t_0=14$, $V/t_0=3$, $A_t/t_0=0.10$,
$\omega_d=\Delta_{\rm eff}/\hbar$, and $\tau=tt_0/\hbar$;
$\mu=\mu_{12}^{(T)}$ is adjusted at each temperature.}
\label{fig:ideal_charging}
\end{figure*}

Figure~\ref{fig:ideal_charging}(a) is the time-domain realization of the
interaction-dependent selection rule in Eq.~\eqref{eq:Mtun}: the
resonant drive produces an almost complete transfer between the two even
singlets while leaving the spectator manifold unchanged. At the
reference temperature its conditional population is
$P_{\rm sp}^{(2)}\simeq0.002603$ and is barely visible on the scale of panel (a).
Panel (b) then
verifies Eq.~\eqref{eq:W_energy_identity} to numerical precision.  Combining this
identity with the sector-selective barrier transition gives
{
\begin{align}
\Delta E_B(t)&=\mathcal W_{U(1)}(t)=p_2\mathcal W_2(t),\nonumber\\
p_1\mathcal W_1(t)&=p_3\mathcal W_3(t)=0.
\label{eq:ideal_sector_chain}
\end{align}
}

The same noiseless identity holds under the additional spin
constraint. Because the initial Gibbs state is passive within each
$(N,S)$ block and the barrier drive preserves these blocks,
$\Est(t)=\WNS[\rho(t)]=\Wu[\rho(t)]$ when $\delta\epsilon=0$.
Thus the distinction in Fig.~\ref{fig:mixing_loss} arises from
unresolved ensemble mixing rather than from a different ideal
charging channel.

At $\mu_{12}^{(T)}$, the sector probabilities satisfy $p_1=p_2=1/2$, yet only the $N=2$ sector
contributes to the extractable work $\mathcal W_{U(1)}$.  The loading probability of a sector
must therefore be distinguished from its response to the charging drive. Neither quantity
establishes controllability during discharge.  In the low-temperature
limit and for nearly complete inversion of the active doublet, the full-inversion reference is
$\mathcal W_{U(1)}^{\rm max}\simeq p_2\Delta_{\rm eff}\simeq5.85\,t_0$,
consistent with the numerical maximum in Fig.~\ref{fig:ideal_charging}.
Figure~\ref{fig:ideal_charging}(c) separately shows how the full-inversion
reference falls with temperature and compares it with the maximum reached by
the specified pulse over $0\le\tau\le35$. The chemical potential is
recomputed at the corresponding $\mu_{12}^{(T)}$ for each temperature.
The close correspondence in the displayed regime does not establish optimal
control or the attainability of the symmetry-constrained discharge optimum.
This ideal result provides the reference against which the noise analysis below
will be compared.

Thermal occupation of states outside the active doublet limits the work available from ideal
inversion. Equations~\eqref{eq:thermal_spectator_population}
and \eqref{eq:thermal_inversion_bound} show that the charging capacity
is determined jointly by the sector weight $p_2$, the thermal singlet
population imbalance, and the transition energy $\Delta_{\mathrm{eff}}$.
A larger $U_{\mathrm{eff}}$ increases the latter while narrowing the
singlet--triplet gap in Eq.~\eqref{eq:singlet_gap}; therefore, the
transition matrix element alone does not determine the thermally
accessible work.

\subsection{Noise-induced mixing losses and spin-constrained work}
\label{sec:noise_results}

We now calculate how unresolved detuning fluctuations reduce the work extractable under
particle number conservation. The
calculations retain all charge sectors of the extended-Hubbard model.
The initial state is the grand-canonical Gibbs state at
$\mu=\mu_{12}^{(T)}$ and $\epsilon=0$, with
$U/t_0=14$, $V/t_0=3$, and $k_BT/t_0=0.05$. Its preparation weights
satisfy $p_1\simeq p_2\simeq 1/2$, while the other sectors have
negligible thermal populations. The charging field has
$A_t/t_0=0.10$, $\phi=0$, and
$\omega_d=\Delta_{\mathrm{eff}}/\hbar$, with the smooth turn-on
defined in Eq.~\eqref{eq:drive_envelope_numeric}.

Fluctuations at frequencies well below the inverse charging time are approximately constant
during a pulse. We therefore study a Gaussian ensemble of static detuning offsets,
corresponding to the quasistatic limit of low-frequency charge noise. Each realization is
propagated with
$\delta\epsilon_r(t)=\delta\epsilon_r$, where the constant detuning
is drawn from a zero-mean Gaussian distribution of rms width
$\sigma_\epsilon$. Gaussian ensemble averages are evaluated through
Gauss--Hermite quadrature, with the density matrices averaged
\emph{before} the evaluation of constrained ergotropy. The model
prepares the same Gibbs state at zero detuning for every realization and
applies a quasistatic offset during the isolated stroke; it does not model
an offset present during thermal equilibration. If slow disorder is present
already during loading, the initial conditional states and charge weights
can themselves depend on the realization and require a separate average. In particular,
the mean state $\bar\rho(t)$ is generally not related to the initial
state by a single unitary, even though every microscopic trajectory
preserves the full charge-sector decomposition.

\begin{figure}[t]
\centering
\includegraphics[width=0.8\columnwidth]{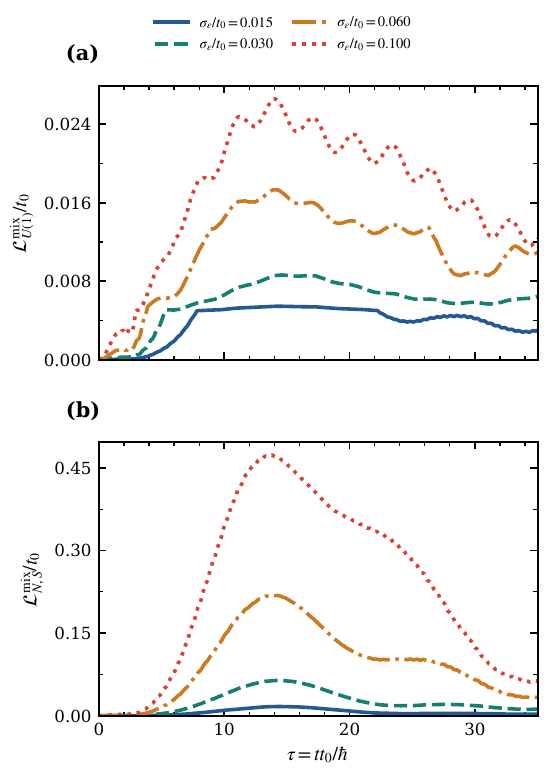}
\caption{\textbf{Increase in passive energy caused by detuning noise under two extraction constraints.}
(a) Mixing loss $\mathcal L_{U(1)}^{\rm mix}/t_0$ when only particle number is conserved.
(b) Mixing loss $\mathcal L_{N,S}^{\rm mix}/t_0$ when particle number
and total spin are both conserved.
The same unobserved, quasistatic Gaussian detuning ensemble is used in
both panels, with rms strengths
$\sigma_\epsilon/t_0=0.015,0.030,0.060,0.100$;
density matrices are averaged before passive energies are computed.
Parameters: $U/t_0=14$, $V/t_0=3$, $k_BT/t_0=0.05$,
$A_t/t_0=0.10$, $\mu=\mu_{12}^{(T)}$,
$\omega_d=\Delta_{\rm eff}/\hbar$, and $\tau=tt_0/\hbar$.
The difference is due to the extraction constraint, not to spin
relaxation or a second noise process.}
\label{fig:mixing_loss}
\end{figure}

Figure~\ref{fig:mixing_loss} gives a direct numerical illustration of
the thermodynamic distinction established in Sec.~\ref{subsec:mixing_loss}.
For the initially passive grand-canonical preparation,
Eq.~\eqref{eq:Lmix} reduces to
$\mathcal L^{\mathrm{mix}}_{U(1)}(t)
=\Delta E_B(t)-\mathcal W_{U(1)}[\bar\rho(t)]$.
The plotted quantity is therefore the increase in sectorwise passive
energy caused by discarding the noise-realization label. It is not
identical to the reduction of stored energy or of ergotropy relative
to the noiseless charging protocol, since fluctuations may also
modify the energy deposited in the battery.

Figure~\ref{fig:mixing_loss}(a) shows the mixing loss when extraction conserves particle
number. Panel (b) adds the requirement of total spin conservation while keeping the charged
state and noise ensemble identical.
At $\tau=17$ and $\sigma_\epsilon/t_0=0.10$, the refined
values obtained with $\delta\tau=0.005$ and $n_q=25$ are
$\Lmix/t_0\simeq0.024652$ and $\LNS/t_0\simeq0.412325$.

The contrast follows from the different passive rearrangements allowed during extraction.
Only the benchmark that conserves charge alone permits the effective spectral
rearrangement between the singlet and triplet subspaces of $N=2$.
At this operating point, the $N=1$ sector contributes approximately
$0.00236\,t_0$ to either mixing loss. The $N=2$ contribution is about
$0.02229\,t_0$ under $U(1)$ conservation but $0.40997\,t_0$ when
spin is also conserved. Thus the entire gap of approximately
$0.38767\,t_0$ between the two totals comes from the half-filled
singlet--triplet energy ordering, not from a second noise process.
The spectral identity \eqref{eq:exact_ST_gap} makes the mechanism
quantitative. At this point, direct diagonalization of the subnormalized singlet block of the
normalized, noise-averaged $N=2$ state gives
$(s_1,s_2,s_3)\simeq(0.92590262,0.04792106,0.02357306)$, while each
triplet has weight $q\simeq0.00086775$. Hence $q<s_3$ and the last
two terms of Eq.~\eqref{eq:exact_ST_gap} yield, after multiplication by
$p_2\simeq1/2$, a work-deficit
$\mathcal G_{\rm ST}/t_0\simeq0.38767255$.
This independently reproduces $\LNS-\Lmix$ and shows that the
large separation between the plotted losses results from the
second and third singlet eigenvalues crossing the triplet spectral
weight, not from different stochastic generators.
The two losses should not be identified with the total reduction
in work relative to noiseless charging, because the noise also
changes the mean stored energy. The large mixing-loss contrast is
also distinct from the fractional difference in constrained work:
at this point $\mathcal G_{\rm ST}/\mathcal W_{U(1)}\simeq0.0694$.
The additional spin restriction reduces $\mathcal W_{U(1)}$ by approximately $6.94\%$. The
larger factor of $16.7$ refers only to the ratio of the mixing losses.

\subsubsection{Perturbative onset of the spin-conservation deficit}
Equation~\eqref{eq:exact_ST_gap} also allows the onset of the deficit to be estimated without
first diagonalizing the averaged state. In the physically
relevant branch $q<s_1$, the first onset occurs when the second
singlet eigenvalue crosses the triplet weight, $s_2=q$. At low
preparation temperature, a perturbative expansion of the driven
singlet propagator with respect to the dimensionless detuning
$\eta=\delta\epsilon/t_0$ gives
\begin{equation}
\begin{aligned}
s_2(\sigma_\epsilon,\tau)
&=\chi(\tau)\left(\frac{\sigma_\epsilon}{t_0}\right)^2
+O\!\left[(\sigma_\epsilon/t_0)^4\right]
+O(\lambda_{\rm exc}),\\
\chi(\tau)&=\lambda_g\sum_{j\ne g}
\left|\langle j|K_\tau|g\rangle\right|^2,\\
K_\tau&=\int_0^\tau\!ds\,
U_0^\dagger(s)Z_{D,2}U_0(s).
\end{aligned}
\label{eq:singlet_noise_susceptibility}
\end{equation}
Here $U_0(s)$ is the driven, noiseless singlet propagator in scaled
time, $Z_{D,2}$ is the singlet detuning operator, $|g\rangle$ is
the initial singlet ground state, $\lambda_g$ its conditional
$N=2$ population, and $\lambda_{\rm exc}$ is the exponentially
small total weight of the two excited singlets. The normalization
of $Z_{D,2}$ is the one multiplying $\delta\epsilon/t_0$ in the
singlet Hamiltonian; no separate factor of $t_0$ is implicit.
The calculation retains the \emph{full time-dependent barrier drive}.
The resulting leading-order estimate is
\begin{equation}
\frac{\sigma_{\epsilon,c}}{t_0}
\simeq\sqrt{\frac{q}{\chi(\tau)}}\,,\qquad \chi(\tau)>0.
\label{eq:predicted_sigma_threshold}
\end{equation}
For $\tau=17$, the thermal conditional triplet weight is
$q\simeq0.000867751$ and $\chi(17)\simeq11.7557$, predicting
$\sigma_{\epsilon,c}/t_0\simeq0.008592$. Direct propagation and
Gauss--Hermite averaging instead bracket the first crossing by
$0.00864<\sigma_{\epsilon,c}/t_0<0.00865$; the relative difference
from the leading estimate is below $1\%$. The weakest detuning shown
in Fig.~\ref{fig:mixing_loss} ($0.015t_0$) is already above this
predicted onset. The crossing is a change in the passive ordering of a finite spectrum, not a
thermodynamic phase transition.
The formula assumes weak quasistatic Gaussian fluctuations,
$U>V$, full spin symmetry, the common zero-detuning Gibbs preparation,
and nearly empty excited singlets. Appendix~\ref{app:noise_onset}
provides the projection-based perturbative derivation and numerical
checks; no uniqueness of crossings at larger noise is claimed.

The mixing loss accounts for only part of the reduction in extractable work relative to
noiseless charging. For an initially passive
preparation, the two effects obey the exact bookkeeping identity
\begin{align}
\label{eq:total_work_loss}
\mathcal W_C^{(0)}-\mathcal W_C^{(\sigma)}
&=\Delta E_B^{(0)}-\Delta E_B^{(\sigma)}
+\mathcal L_C^{\rm mix}(\sigma),\nonumber\\
&\hspace{-2.5mm} C\in\{U(1),(N,S)\}.
\end{align}
Here the superscript $(0)$ denotes the noiseless evolution at the same
charging duration, whereas $(\sigma)$ denotes the unresolved Gaussian
ensemble. Table~\ref{tab:full_work_balance} separates these contributions
at the reference point. The energy decrease of $0.227019\,t_0$ combines
with passive-energy increases of $0.024652\,t_0$ and $0.412325\,t_0$ to
yield total work reductions of $0.251671\,t_0$ and $0.639344\,t_0$,
respectively. Thus the factor-of-$16.7$ contrast applies to the
\emph{mixing contributions}, not to the total work reductions.
\begin{table}[t]
\caption{\textbf{Stored energy, constrained work, and mixing losses at the reference point.}
Noiseless and quasistatic-noise calculations use the same charging pulse
at $\tau=17$, $U/t_0=14$, $V/t_0=3$, $A_t/t_0=0.10$,
$k_BT/t_0=0.05$, and $\mu=\mu_{12}^{(T)}$.
The noisy ensemble has $\sigma_\epsilon/t_0=0.10$ and is computed
with $\delta\tau=0.005$ and $n_q=25$ Gauss--Hermite nodes.
All tabulated energy, work, and loss values are in units of $t_0$.}
\label{tab:full_work_balance}
\begin{ruledtabular}
\begin{tabular}{lcc}
Quantity & Noiseless & Quasistatic noise \\
\hline
$\Delta E_B$ & 5.835216 & 5.608197 \\
$\mathcal W_{U(1)}$ & 5.835216 & 5.583544 \\
$\mathcal W_{N,S}$ & 5.835216 & 5.195872 \\
$\mathcal L^{\mathrm{mix}}_{U(1)}$ & 0 & 0.024652 \\
$\mathcal L^{\mathrm{mix}}_{N,S}$ & 0 & 0.412325 \\
\end{tabular}
\end{ruledtabular}
\end{table}

All four curves in each panel are nonnegative, consistent with the convexity result in
Eq.~\eqref{eq:noise_inequality}. At comparable
charging times, stronger detuning disorder generally produces a
larger mixing-induced loss in the regime investigated. The temporal
response is nevertheless nonmonotonic: the loss increases during the
initial part of the driven evolution and subsequently exhibits
oscillatory variations. Such behavior is compatible with the coherent,
finite-duration charging dynamics combined with ensemble averaging;
it should not be interpreted as irreversible dissipation along
individual trajectories. Neither an exponential-decay law nor a
monotonic increase at every instant follows from the random-unitary
model.

Importantly, the plotted loss includes every occupied charge sector,
not only the resonantly driven even-singlet doublet. Although the
barrier pulse by itself leaves the $N=1$ sector dynamically inactive
at $\epsilon=0$, the detuning perturbation does not commute with its
static hopping Hamiltonian. The sector weights remain constant while
the averaged conditional states can become more passive. In particular,
the weak-noise calculation in Appendix~\ref{app:single_electron} identifies
a filter-function-like $N=1$ contribution that oscillates with the orbital
splitting and retains the applied barrier waveform. Its modulation explains
part of the fine structure in Fig.~\ref{fig:mixing_loss}(a), without
contradicting the dark-sector rule for ideal barrier-only charging.
Thus the separation between energy injection and usable work emerges entirely
within a particle-number-conserving description, without invoking
particle leakage or a phenomenological relaxation channel. The optimum $\mathcal W_{U(1)}$
remains an upper bound on extraction when particle number is conserved; imposing total spin
conservation gives the tighter bound $\mathcal W_{N,S}$.

\paragraph{Numerical procedure and convergence.}
\label{sec:numerics}

The temperature--noise map reuses the temperature-independent driven
propagators across Gibbs preparations, and calculates the passive-ordering
difference from eigenvalues of the density matrix averaged over Gaussian
offsets. Checks of the thermal boundary at larger noise amplitudes use increased
Gaussian quadrature orders and direct passive spectral sorting. The charge blocks are propagated by unitary midpoint
matrix-exponential steps using the scaled time $\tau=tt_0/\hbar$,
$\Delta\tau=0.01$, and $0\le\tau\le35$. Sector probabilities are
preserved for each realization, and the ensemble average is taken
only after the individual density matrices have been propagated.
The noiseless energy--ergotropy identities, positivity of the averaged states,
and nonnegativity of both mixing losses provide independent
numerical consistency checks. Quantitative convergence should be assessed by
refining the time step and the Gaussian quadrature independently
before interpreting the fine oscillatory structure of Fig.~\ref{fig:mixing_loss}.
An independent check of the $s_2-q$ crossing is presented in
Appendix~\ref{app:parity_convergence}. The charging protocol is
held fixed throughout these refinements. The numerical propagation and Gaussian
averaging procedure is detailed in Appendix~\ref{app:checks}, while separate
time-step and quadrature refinements of the thermal crossover are documented
in Appendix~\ref{app:parity_convergence}.

The auxiliary interaction--noise scan was evaluated at
$U_{\rm eff}/t_0=5,11,15$ using a Python implementation with
$\delta\tau=0.01$ and $n_q=25$. At the reference interaction
$U_{\rm eff}/t_0=11$, the main result was checked by joint
refinement from $(\delta\tau,n_q)=(0.01,17)$ to $(0.005,25)$,
as tabulated in Appendix~\ref{app:checks}. The corresponding changes
are approximately $4.50\times10^{-4}\,t_0$ in $\mathcal W_{N,S}$
and $3.66\times10^{-5}\,t_0$ in
$\mathcal L^{\rm mix}_{N,S}$. These tests check numerical
consistency at the reference point; convergence across the full
interaction scan has not been established.
The low-frequency $1/f$ description in Eq.~\eqref{eq:PSD} motivates
the quasistatic approximation, but Fig.~\ref{fig:mixing_loss} displays
Gaussian, time-independent noise realizations: it is not a simulation of a $1/f$ process
resolved in time. For a spectrum restricted
to $1\,\mathrm{Hz}\le f\le1\,\mathrm{MHz}$, even the highest frequency
changes negligibly during a roughly $136$-ps charging window.
Consequently, the present figure establishes the mixing-loss behavior
for Gaussian quasistatic detuning disorder; identifying dynamical
spectral effects would require a separately characterized higher-frequency
noise band.

\subsection{Temperature--noise crossover of the spin-work deficit}
\label{sec:temperature_crossover}

At low temperature, unresolved detuning fluctuations produce different mixing losses depending
on whether extraction must conserve total spin in addition to particle number. We now identify
how the thermal triplet occupation controls this
contrast, keeping the Hamiltonian, pulse, and noise protocol fixed while varying
the preparation temperature. The thermal analysis follows from the spectral criterion derived
above and does not require a phenomenological decay law.
At each temperature we choose $\mu=\mu_{12}^{(T)}$, so that the
$N=1$ and $N=2$ loading weights are equal by construction. The
remaining charge-sector probabilities are still obtained from the full
grand-canonical partition function; they are not artificially set to
zero. This choice maintains equal loading probabilities for $N=1$ and $N=2$ as temperature varies.

For the spin-invariant averaged $N=2$ state, let
$s_1\ge s_2\ge s_3$ denote the singlet eigenvalues and
$q$ the population of any one of the three degenerate triplets.
Equation~\eqref{eq:exact_ST_gap} yields the exact deficit
$\mathcal G_{\rm ST}=\mathcal W_{U(1)}-\mathcal W_{N,S}$.
The zero-deficit region satisfies $s_2\le q\le s_1$ when $p_2>0$.
Along the branch realized by the reference simulations, $q<s_1$, and the
first onset or disappearance of the additional spin constraint occurs at
\begin{equation}
  s_2(T,\sigma_\epsilon;\tau)=q(T),
  \label{eq:thermal_spectral_boundary}
\end{equation}
where $s_2$ depends on the noise-averaged singlet dynamics and $q(T)$
is fixed by the initial Gibbs preparation because the triplet block is
invariant under the spin-independent Hamiltonian. At other points, such
as a hypothetical branch with $q>s_1$, the complete positive-part
expression in Eq.~\eqref{eq:exact_ST_gap} must be used. The equality identifies
a change in the passive ordering of a finite spectrum, rather than a quantum
phase boundary of the kind studied using Wigner-function diagnostics in
interacting spin chains \cite{ElHawary2024PhaseSpace}. No thermodynamic phase
transition is implied.

At fixed charging time $\tau=17$, we evaluate both constrained ergotropies
from the same ensemble-averaged conditional density matrices on a
$(k_BT/t_0,\sigma_\epsilon/t_0)$ grid. Each detuning realization is drawn
from the same zero-mean, quasistatic Gaussian distribution as for
Fig.~\ref{fig:mixing_loss}. Since the Hamiltonian and prescribed noise
statistics do not depend on the preparation temperature, the numerical
propagators can be reused across temperatures; only the initial Gibbs
states and grand-canonical weights are recomputed. This removes avoidable
propagation differences from the thermal comparison.

\begin{figure*}[t]
\centering
\includegraphics[width=0.98\textwidth]{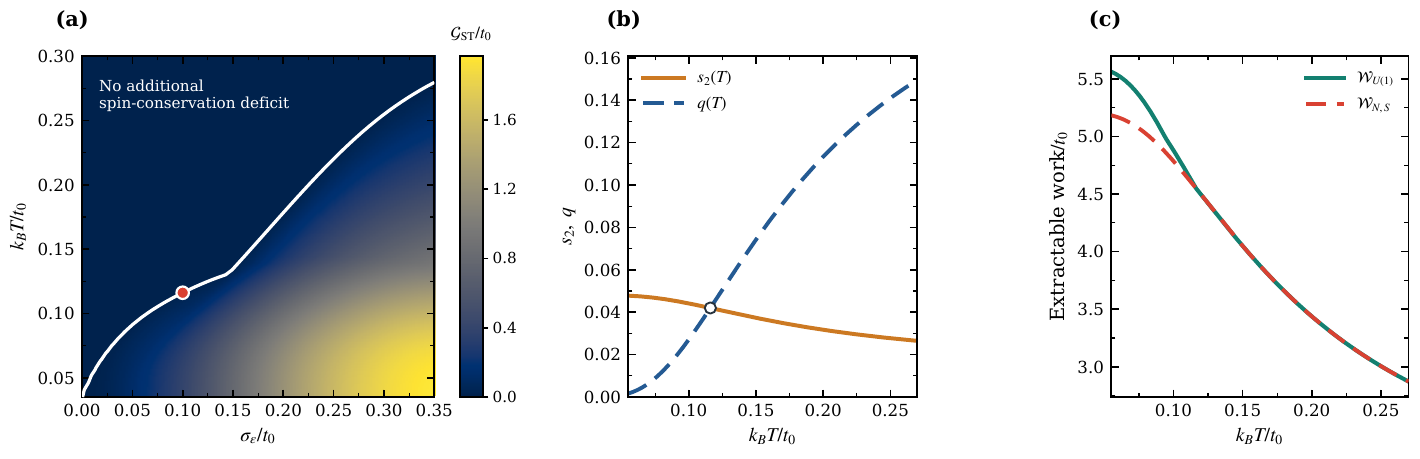}
\caption{\textbf{Thermal crossover of the spin work deficit under detuning noise.}
(a) $\mathcal G_{\rm ST}/t_0=\bigl(\mathcal W_{U(1)}-\mathcal W_{N,S}\bigr)/t_0$
versus $k_BT/t_0$ and quasistatic detuning-noise amplitude
$\sigma_\epsilon/t_0$, at fixed $\tau=17$.
The solid boundary follows the numerically determined condition $s_2=q$
on the branch $q<s_1$; the full positive-part expression is used
at all grid points. The red marker identifies the reference crossover at
$\sigma_\epsilon/t_0=0.10$, examined in panels (b) and (c).
(b) Second singlet eigenvalue $s_2$ and population $q$ of each triplet as functions of temperature at $\sigma_\epsilon/t_0=0.10$; their crossing occurs
at $k_BT_c/t_0\simeq0.11607$.
(c) Corresponding constrained ergotropies of the same averaged state.
Parameters: $U/t_0=14$, $V/t_0=3$, $A_t/t_0=0.10$,
$\omega_d=\Delta_{\rm eff}/\hbar$; $\mu=\mu_{12}^{(T)}$ is
recalculated at each temperature. The boundary marks a change in passive spectral ordering, not a phase transition.}
\label{fig:thermal_noise_crossover}
\end{figure*}

Figure~\ref{fig:thermal_noise_crossover}(a) maps the nonnegative additional
extraction restriction. Its boundary is determined by direct spectral
sorting, with the curve $s_2=q$ overlaid only on the branch $q<s_1$.
Panel (b) resolves the underlying eigenvalue crossing at
$\sigma_\epsilon/t_0=0.10$:
for $U/t_0=14$, $V/t_0=3$, $A_t/t_0=0.10$, and $\tau=17$,
we obtain $k_BT_c/t_0\simeq0.11607$.
In the representative static energy normalization $t_0=0.17\,\mathrm{meV}$,
this corresponds to approximately $0.229\,\mathrm{K}$. The reference
preparation temperatures $k_BT/t_0=0.05$ and $0.25$ correspond to
approximately $99$ and $493\,\mathrm{mK}$ for this same energy scale.
The values follow from the illustrative energy scale and do not represent measured electron
temperatures. Neither temperature is a
universal critical value: the crossing depends on the drive duration,
interaction parameters, and selected detuning-noise ensemble.

The thermal dependence can be separated from the singlet dynamics.
Let $|g\rangle=|E_-^{(2)}\rangle$ denote the initial singlet ground
state, and define
\begin{equation}
 A_g(\sigma_\epsilon,\tau)=\mathbb E_\eta\!\left[
 U_\eta(\tau)|g\rangle\langle g|U_\eta^\dagger(\tau)\right],
 \label{eq:thermal_Ag}
\end{equation}
where $U_\eta$ is the singlet propagator of
Appendix~\ref{app:noise_onset} and
$\eta\sim\mathcal N(0,(\sigma_\epsilon/t_0)^2)$.
The ordered eigenvalues $a_1\ge a_2\ge a_3\ge0$ of the
unit-trace matrix $A_g$ satisfy $\sum_i a_i=1$. They do not depend
on preparation temperature when the pulse, charging time, and noise
ensemble are fixed. The initial singlet-ground-state weight and
each triplet weight are
\begin{equation}
 \lambda_g(T)=\frac{e^{-\beta E_-^{(2)}}}{Z_2},\qquad
 q(T)=\lambda_g(T)e^{-J_{\rm ST}/(k_BT)},
 \label{eq:thermal_weights}
\end{equation}
where $J_{\rm ST}=V-E_-^{(2)}$. Neglecting the initial excited-singlet
populations gives $s_i(T,\sigma_\epsilon;\tau)\simeq
\lambda_g(T)a_i(\sigma_\epsilon,\tau)$, with an error bounded in
Appendix~\ref{app:thermal_onsets}. On the branch $q<s_1$ of
Fig.~\ref{fig:thermal_noise_crossover}, the boundary $s_2=q$
therefore satisfies the semi-analytic relation
\begin{equation}
 k_BT_c\simeq\frac{J_{\rm ST}}{\ln[1/a_2(\sigma_\epsilon,\tau)]}.
 \label{eq:semi_analytic_Tc}
\end{equation}
This relation requires $a_2>0$ and a small thermal correction relative
to $a_2$ at the crossing. If $a_2=0$, this approximation does not
predict a positive finite crossover temperature. The general
zero-deficit criterion remains $s_2\le q\le s_1$, or approximately
$a_2\le\exp[-J_{\rm ST}/(k_BT)]\le a_1$; hence no general statement
about all higher temperatures follows from the single boundary.
At $\tau=17$ and $\sigma_\epsilon/t_0=0.10$,
Eq.~\eqref{eq:semi_analytic_Tc} gives $k_BT_c/t_0\simeq0.11607$,
consistent with Table~\ref{tab:thermal_noise_thresholds}.

The crossover thus compares a Boltzmann factor controlled by the
singlet--triplet gap with a spectrum set by the charging pulse and
noise ensemble. Charging does not require transitions between singlets and triplets. The
loading factor $p_2$ cancels from the zero-deficit
condition for $p_2>0$, although it remains in the work values.

An additional structure in the
$\sigma_\epsilon$ dependence follows from the exact parity symmetry
of the averaged state, Eq.~\eqref{eq:averaged_parity}. At fixed
$k_BT/t_0=0.12$, the smaller even-singlet eigenvalue lies below
the odd-parity eigenvalue at $\sigma_\epsilon/t_0=0.145$, but
above it at $0.150$. Along the computed thermal boundary,
$\lambda_{{\rm even},-}-\lambda_{\rm odd}$ likewise changes from
approximately $-3.54\times10^{-4}$ at $0.145$ to
$+3.17\times10^{-3}$ at $0.150$ (using $\delta\tau=0.005$). The second-largest singlet
eigenvalue $s_2$ therefore switches its parity-sector origin
in this interval. The exchange in eigenvalue order accounts for the change in slope of the
$s_2=q$ contour in Fig.~\ref{fig:thermal_noise_crossover}(a).
The parity crossing changes which singlet eigenvalue enters $s_2$. The condition $s_2=q$
instead determines when conservation of spin ceases to reduce the extraction optimum.
The corresponding parity-resolved values are listed in
Table~\ref{tab:parity_branch_crossing} of Appendix H.
The numerical bracket is protocol specific; no thermodynamic
phase transition or universality is implied.

The disappearance of $\mathcal G_{\rm ST}$ upon warming must be distinguished
from an increase in extractable work. In Fig.~\ref{fig:thermal_noise_crossover}(c)
both ergotropies decline over the displayed thermal window, even when
$\mathcal W_{U(1)}=\mathcal W_{N,S}$. The conserved spin
symmetry and its admissible unitary restriction remain in place:
only their additional energetic cost becomes zero for these states. For the reference noise
amplitude, the triplet weight increases with
temperature while the second singlet eigenvalue decreases across the
plotted crossing region;
the resulting passive ordering eliminates the \emph{additional} restriction
imposed by spin conservation. Mixing-induced passive-energy loss can remain
positive under both symmetry classes.

\begin{table*}[t]
\caption{\textbf{Temperature dependence of stored energy, constrained work, and mixing losses.}
All energy and work entries are in units of $t_0$ at $\tau=17$,
$U/t_0=14$, $V/t_0=3$, and $A_t/t_0=0.10$;
$\mu=\mu_{12}^{(T)}$ is set at each temperature.
Superscripts $(0)$ and $(\sigma)$ denote respectively the noiseless pulse
and the ensemble-averaged evolution with quasistatic Gaussian detuning
$\sigma_\epsilon/t_0=0.10$.
The spin-work deficit is $\mathcal G_{\rm ST}=\mathcal W_{U(1)}-\mathcal W_{N,S}$
for the same averaged state.}
\label{tab:thermal_work_comparison}
\begin{ruledtabular}
\begin{tabular}{cccccccc}
$k_BT/t_0$ & $\Delta E_B^{(0)}$ & $\Delta E_B^{(\sigma)}$ &
$\mathcal W_{U(1)}^{(\sigma)}$ & $\mathcal W_{N,S}^{(\sigma)}$ &
$\mathcal L_{U(1)}^{\rm mix}$ & $\mathcal L_{N,S}^{\rm mix}$ & $\mathcal G_{\rm ST}$ \\
0.05 & 5.835216 & 5.608197 & 5.583544 & 5.195872 & 0.024652 & 0.412325 & 0.387673 \\
0.25 & 3.376160 & 3.245803 & 3.006246 & 3.006246 & 0.239557 & 0.239557 & 0.000000
\end{tabular}
\end{ruledtabular}
\end{table*}

Table~\ref{tab:thermal_work_comparison} illustrates the distinction
between the two mechanisms. At $k_BT/t_0=0.05$ the additional
spin-constrained deficit is $0.387673\,t_0$, whereas it vanishes at
$0.25$ for the same $\sigma_\epsilon/t_0=0.10$ and $\tau=17$.
Nevertheless, the charge-constrained mixing loss \emph{increases} from
$0.024652\,t_0$ to $0.239557\,t_0$. The disappearance of the additional deficit reflects a
change in passive population ordering; it does not indicate protection of stored work against
thermal effects.
Additional noise-dependent crossover temperatures are listed in
Table~\ref{tab:thermal_noise_thresholds} of the Appendix. They are
numerical results for this specified protocol, not a monotonicity
law. The crossing location requires careful Gaussian-quadrature
refinement at larger noise amplitudes. The underlying methodology and
convergence checks are documented in the Appendices; supporting numerical
data and scripts are available from the corresponding author upon reasonable
request.
The perturbative low-noise estimate in Eq.~\eqref{eq:predicted_sigma_threshold}
is not extrapolated to the high-noise part of this map, where its
small-noise assumptions need not hold. The stronger-noise boundary
is determined by the sorted eigenvalues of the fully averaged
state, including the parity-branch exchange discussed above.
\subsection{Physical implications, relation to previous work, and limitations}
\label{sec:discussion}
The results distinguish the roles of preparation, charging dynamics, and extraction.
Grand-canonical loading sets the weights of the five electron-number sectors,
barrier modulation determines which within-sector transitions can be driven,
and conservation laws define the unitary class against which stored work is
benchmarked. For the same prepared and averaged state, the two extraction optima differ only
because their allowed population rearrangements differ. The additional deficit does not imply
spin relaxation. In particular, the thermal crossover
of Fig.~\ref{fig:thermal_noise_crossover} only removes the \emph{additional}
spin restriction; it does not restore work lost through unresolved ensemble
mixing or increase the charging capacity.

Finite
$U_{\rm eff}=U-V$ activates the even-singlet barrier transition, whereas the
same relative interaction controls the singlet--triplet thermal gap and the
second-order susceptibility of the driven gap to detuning.
The comparison of barrier matrix elements, thermal populations, and curvature
in Figs.~\ref{fig:interaction_selectivity} and \ref{fig:ideal_charging}
therefore identifies competing microscopic trends rather than a universally
optimal interaction. A supplementary three-point scan at
$U_{\rm eff}/t_0=5,11,15$, with $V/t_0=3$, $A_t/t_0=0.10$,
$k_BT/t_0=0.05$, $\tau=17$, and $\sigma_\epsilon/t_0=0.10$,
shows that the ergotropy under the spin constraint can grow while the ratio to its
noiseless value falls. Specifically, the two endpoint values are approximately
$2.812\,t_0$ and $6.146\,t_0$ for the work, and $0.956$ and $0.805$
for the retention ratio. The driving frequency and chemical potential are
recalculated at each interaction. These discrete points do not establish a
monotonicity theorem or a device optimum; convergence across this entire scan
remains to be demonstrated.

The dependence on temperature extends studies of charging and filling in Hubbard quantum batteries
\cite{Konar2022}. Our comparison of passive energies uses the framework of Mitsuhashi \emph{et al.} for work extraction under symmetry constraints
\cite{Mitsuhashi2022}, applied here to a concrete finite singlet--triplet
spectrum. The resulting crossover is not the activation of total ergotropy
reported in dissipatively charged spin-battery models
\cite{Yang2026Thermal}: in our case, both ergotropies can remain positive on
both sides of $s_2=q$, and only their difference vanishes. We make no claim
that the charge-sector factorization is a new general theorem, or that a
specified finite set of electrical discharge gates attains either symmetry
benchmark. The distinction between formal optimal extraction and a designated
battery--load coupling is also emphasized in Hubbard--Holstein battery work
\cite{Ostberg2026}.

The parameter choice is guided by experimental energy scales but is not fitted to a specific
double quantum dot. With $t_0\simeq0.17\,{\rm meV}$ and
$U_{\rm eff}/t_0=11$, the direct one-photon even-singlet transition would
require barrier modulation near $0.48\,{\rm THz}$. The representative
Hubbard parameter set does not establish that this pulse is technologically
available. The calculation also neglects higher orbitals, spin--orbit
coupling, phonon relaxation, and device-specific cross-capacitance effects.
The stochastic curves describe \emph{quasistatic Gaussian} detuning offsets,
not a broadband $1/f$ trajectory resolved in time. The exact symmetry results apply to the
Hamiltonian studied here; quantitative predictions for a particular device would require the
omitted physical processes and a calibrated drive.

\section{Conclusion}
\label{sec:conclusion}

We have examined charging and work extraction in an interacting Hubbard
double quantum dot with reservoir preparation over all electron number
sectors. Conservation of particle number allows the extraction optimum to be
calculated separately in each sector. At zero detuning, modulation of the
symmetric tunnel barrier drives only the two even singlets at half filling;
the remaining states are spectators under the ideal pulse. The matrix element
for this transition vanishes when $U=V$, despite the finite singlet splitting
$\Delta_{\mathrm{eff}}=4t_0$. The relative interaction $U-V$ thus controls
the availability of the charging transition.

For each realization of quasistatic detuning disorder, the evolution remains
unitary and conserves both particle number and total spin. Averaging over
unresolved realizations nevertheless raises the minimum passive energy
relative to its value along individual trajectories. This mixing loss is
distinct from changes in the mean energy deposited during charging. When extraction is required
to conserve spin as well as particle
number, the additional loss is determined entirely by the singlet and triplet
spectrum of the $N=2$ sector. Passive spectral sorting gives an exact formula
for the difference between the two extraction optima. At low temperature, the
noise amplitude for the onset of this difference is predicted by a perturbative
response calculation and agrees with direct numerical propagation.

Increasing the preparation temperature changes the triplet occupation and
can remove the additional work deficit without increasing either extraction
optimum. For the reference protocol, the crossover occurs at
$k_BT_c/t_0\simeq0.11607$ for $\sigma_\epsilon/t_0=0.10$. Its boundary is
controlled by the equality $s_2=q$ between a singlet eigenvalue and the
population of an individual triplet. The temperature dependence can be
expressed in terms of the initial Gibbs weights and the singlet state produced
by the same charging pulse, with a bound on the effect of initially occupied
excited singlets. The symmetric noise ensemble also restores parity, allowing
an exchange in the ordering of singlet eigenvalues from different parity
sectors. This accounts for the change of branch along the thermal boundary.
A separate interaction scan illustrates that a larger absolute work output
may be accompanied by poorer fractional retention under noise. These results
refer to extraction optima defined by conservation laws. They do not establish
that the same work can be recovered using a finite sequence of electrical
control gates.

\begin{acknowledgments}
K. Loukhssami gratefully acknowledges financial support from CNRST Morocco
through the PhD Associate Scholarship PASS initiative.
C. Shang was supported by the RIKEN Hakubi Fellows Program.
\end{acknowledgments}

\appendix
\section{Thermal occupation and the active singlet fraction}
\label{app:thermal}

At symmetric detuning, the six conditional energies of $N=2$ are
$E_-^{(2)}$, the threefold-degenerate $V$, the odd-singlet energy $U$,
and $E_+^{(2)}$. Writing
$J_{\rm ST}=V-E_-^{(2)}=(\Delta_{\rm eff}-U_{\rm eff})/2$,
the partition function may equivalently be factored as
\begin{equation}
Z_2=e^{-\beta E_-^{(2)}}\left[
1+3e^{-\beta J_{\rm ST}}
+e^{-\beta(U-E_-^{(2)})}
+e^{-\beta\Delta_{\rm eff}}\right].
\label{eq:appZ2}
\end{equation}
Thus the conditional fraction of thermally occupied spectator states
follows directly from Eq.~\eqref{eq:thermal_spectator_population}.
For $U/t_0=14$ and $V/t_0=3$ the values below are obtained from the
exact Gibbs probabilities. The last column reports the ideal inversion
work with the charge-sector weight held at $p_2=1/2$, isolating the
within-sector thermal effect from the loading statistics.
\begin{table}[!ht]
\caption{\textbf{Thermal occupation of spectator states and ideal inversion work in the $N=2$ sector.}
The conditional spectator probability $P_{\rm sp}^{(2)}$ is dimensionless.
The ideal inversion work $W_{\rm inv}^{\max}$ is given in units of $t_0$
with the charge-sector loading weight fixed to $p_2=1/2$ only for
that work column. Parameters: $U/t_0=14$, $V/t_0=3$,
$\epsilon=0$.}
\label{tab:appthermal}
\begin{ruledtabular}
\begin{tabular}{ccc}
$k_BT/t_0$ & $P_{\rm sp}^{(2)}$ & $W_{\rm inv}^{\max}/t_0$\\
0.05 & 0.002603 & 5.8371\\
0.10 & 0.081294 & 5.3766\\
0.20 & 0.340035 & 3.8623\\
\end{tabular}
\end{ruledtabular}
\end{table}
These results provide an exact thermal interpretation of
Eq.~\eqref{eq:thermal_inversion_bound}; the comparison in Fig.~\ref{fig:ideal_charging}(c) evaluates the
corresponding temperature-dependent charging capacity; the temperature--noise
crossover is analyzed separately in Sec.~\ref{sec:temperature_crossover}.

\section{Stochastic averaging and numerical verification}
\label{app:checks}

In each charge sector, the Schr\"odinger propagation uses a midpoint
approximation with a unitary matrix exponential. In scaled time
$\tau=tt_0/\hbar$, the update for realization $r$ is
\begin{equation*}
U_{r,N}(\tau+\delta\tau)
=e^{-i\delta\tau H_{r,N}(\tau+\delta\tau/2)/t_0}
U_{r,N}(\tau).
\end{equation*}
We propagate the initial Gibbs state in each charge sector for each static detuning offset and
then average the resulting density matrices.
For $\delta\epsilon\sim\mathcal N(0,\sigma_\epsilon^2)$, an $n_q$-point
probabilists' Gauss--Hermite rule approximates any smooth ensemble observable
$F$ as
\begin{equation*}
\mathbb E_{\delta\epsilon}[F(\delta\epsilon)]
\simeq\frac{1}{\sqrt{2\pi}}\sum_{j=1}^{n_q}
w_j F(\sigma_\epsilon x_j),
\end{equation*}
where $(x_j,w_j)$ are nodes and weights for the weight
$e^{-x^2/2}$. The sector-resolved density matrices are averaged first;
the nonlinear ergotropy and passive energies are evaluated only afterward.
The evolution conserves $\Tr\rho$, positivity and $p_N$ up to
floating-point precision, and verifies
$\Wu=\Delta E_B$ in the ideal case.

At the representative point $U_{\rm eff}/t_0=11$,
$\sigma_\epsilon/t_0=0.10$ and $\tau=17$, the joint refinement of
step size $\delta\tau$ and quadrature order $n_q$ yields
\begin{table}[!ht]
\caption{\textbf{Joint time-step and quadrature refinement of charge-constrained work and mixing loss.}
Values are computed at $\tau=17$, $U_{\rm eff}/t_0=11$,
$V/t_0=3$, $k_BT/t_0=0.05$, $A_t/t_0=0.10$,
$\mu=\mu_{12}^{(T)}$, and $\sigma_\epsilon/t_0=0.10$.
Here $\delta\tau$ is the scaled propagation step and $n_q$ is the
number of Gauss--Hermite nodes. The last two columns are dimensionless because the corresponding
quantities are divided by $t_0$; this table reports joint rather than separate refinements.}
\label{tab:appconvergence}
\begin{ruledtabular}
\begin{tabular}{cccc}
$\delta\tau$ & $n_q$ & $\mathcal W_{U(1)}/t_0$ & $\mathcal L^{\rm mix}_{U(1)}/t_0$\\
0.040 & 11 & 5.573917 & 0.024672\\
0.020 & 13 & 5.581429 & 0.024658\\
0.010 & 17 & 5.583130 & 0.024653\\
0.005 & 25 & 5.583544 & 0.024652\\
\end{tabular}
\end{ruledtabular}
\end{table}

At the finest reported discretization,
$(\delta\tau,n_q)=(0.005,25)$, the spin-preserving values at
$\tau=17$ and $\sigma_\epsilon/t_0=0.10$ are
$\WNS/t_0=5.195872$ and $\LNS/t_0=0.412325$.
The charge-only values are $\Wu/t_0=5.583544$ and
$\Lmix/t_0=0.024652$. The ensemble-averaged state is
identical in both calculations; only the minimization of passive
energy changes.

The mixing loss is substantially less sensitive to joint numerical
refinement than the full oscillatory work signal at this point. The
numbers are a joint check, not independent error estimates for the
quadrature and propagation algorithms.

\section{Weak-noise origin of single-electron oscillations}
\label{app:single_electron}

The ideal barrier modulation is unable to excite the one-electron
sector at zero detuning because the drive commutes with its static
Hamiltonian. A static detuning offset, however, does not commute with
the hopping operator. The small mixing-loss oscillations of the
$N=1$ sector can be obtained perturbatively while retaining the full
barrier waveform. Writing $x=\delta\epsilon/t_0$, the conditional orbital
Hamiltonian in scaled time is
\begin{equation}
\begin{aligned}
 H_1(\tau)/t_0&=-a(\tau)\sigma_x+\frac{x}{2}\sigma_z,\\
 a(\tau)&=1+\frac{A_t}{t_0}g(\tau)
 \cos\!\left(\frac{\Delta_{\mathrm{eff}}}{t_0}\tau+\phi\right).
\end{aligned}
\label{eq:n1_hamiltonian}
\end{equation}
The initial orbital Gibbs state is
$\rho_{1,0}=\bigl(\mathbb I+m\sigma_x\bigr)/2$,
where $m=\tanh(\beta t_0)$. In the interaction picture of the
barrier-only evolution, the transverse detuning perturbation rotates
with an angle $2\Theta(\tau)$, where
\begin{equation}
 \Theta(\tau)=\int_0^\tau a(s)\,ds,\qquad
 F(\tau)=\int_0^\tau e^{2i\Theta(s)}\,ds.
 \label{eq:n1_filter}
\end{equation}
For a zero-mean quasistatic Gaussian offset of variance
$\sigma_\epsilon^2$, the linear transverse response vanishes upon
averaging. Expanding the averaged Bloch-vector length to second order
in $\sigma_\epsilon/t_0$ gives the charge-sector contribution
\begin{equation}
\mathcal L_{1}^{\mathrm{mix}}(\tau)=
 p_1\,\frac{\sigma_\epsilon^2}{2t_0}\,
 \tanh(\beta t_0)\,|F(\tau)|^2
 +O\!\left(\frac{\sigma_\epsilon^4}{t_0^3}\right).
 \label{eq:n1_oscillation}
\end{equation}
This expansion is at fixed finite time and assumes a symmetric
zero-mean distribution. For vanishing ac modulation, $a(\tau)=1$ and
$|F(\tau)|^2=\sin^2\tau$, recovering the orbital period $\pi$
in scaled time. The full drive changes the precise oscillation pattern
through $\Theta$ but not the physical mechanism. At $\tau=17$ and
$\sigma_\epsilon/t_0=0.10$, the second-order expression predicts
approximately $0.00231\,t_0$, compared with about
$0.00236\,t_0$ from numerical propagation of the one-electron block.
It is therefore a controlled weak-noise explanation of the fine
structure, not a replacement for the complete charge-resolved simulation.

\section{Spin-resolved passive energy}
\label{app:spin}

The joint projectors $P_{N,S}$ diagonalize the commuting observables
$\Ne$ and $\mathbf S^2$. Writing the energies within a joint block
as $E_{1|N,S}\le\cdots\le E_{d_{N,S}|N,S}$ and the
eigenvalues of its normalized state in nonincreasing order as
$q_{1|N,S}\ge\cdots\ge q_{d_{N,S}|N,S}$, one obtains
\begin{equation}
\EpNS(\rho)=\sum_{N,S}p_{N,S}
\sum_{j=1}^{d_{N,S}}q_{j|N,S}E_{j|N,S},
\label{eq:app_spin_pass}
\end{equation}
so $\WNS(\rho)=\Tr(H_B\rho)-\EpNS(\rho)$.
For the thermal and driven states used here, every density matrix
commutes with the joint projectors, and the formula applies directly.

The only nontrivial difference between the charge-only and $(N,S)$
passive energies occurs in $N=2$. There the static Hamiltonian is
the direct sum of a three-dimensional singlet Hamiltonian
$H_{2,S=0}$ with energies $E_-^{(2)},U,E_+^{(2)}$ and a
three-dimensional triplet Hamiltonian $VI_3$. A spin-preserving
extraction independently rearranges the three singlet eigenvalues,
while the triplets cannot lower their energy because they are
degenerate. By contrast, $\EpU$ sorts all six eigenvalues
against the joint six-level energy spectrum of $N=2$.
The gap $V-E_-^{(2)}=(\Delta_{\rm eff}-\Ueff)/2$
then provides a low-lying triplet energy that is available to the
charge-only optimization but not to a spin-preserving one.

Because each spin block is preserved along individual noise
realizations, its spectrum is constant. Convexity of the block
ergotropy therefore implies
$\LNS(t)\ge0$ for every ensemble-averaged state.
The additional inequality $\LNS(t)\ge\Lmix(t)$ follows
for the initially passive thermal preparation, as stated in
Eq.~\eqref{eq:spin_loss_difference}. In general, the ordering
of the two mixing losses requires attention to the initial
benchmark values; the ordering of the ergotropies themselves
holds for every state.

\paragraph{Full spin-rotation invariance for the prepared states.}
The microscopic Hamiltonian and both electrical perturbations
commute with each global spin component $S_\alpha$,
$\alpha=x,y,z$. The grand-canonical Gibbs state is therefore
invariant under global $SU(2)$ rotations, as are the propagated
states $\rho_r(t)$ and the ensemble-averaged state $\bar\rho(t)$.
Decomposing each joint charge--spin sector into multiplicity and
spin-representation spaces gives
\begin{equation}
\begin{aligned}
\mathcal H_{N,S}
&\simeq\mathcal M_{N,S}\otimes\mathbb C^{2S+1},\\
H_{N,S}&=h_{N,S}\otimes\mathbb I_{2S+1},\\
\rho_{N,S}&=r_{N,S}\otimes
\frac{\mathbb I_{2S+1}}{2S+1},
\end{aligned}
\label{eq:SU2_schur}
\end{equation}
where $r_{N,S}$ is a normalized state on the multiplicity
space. This structure follows from invariance under all spin
rotations and Schur's lemma; it is not assumed for arbitrary
noninvariant input states \cite{FultonHarris1991,Bartlett2007}.

Let $q_1\ge\cdots\ge q_m$ be the eigenvalues of $r_{N,S}$,
and $e_1\le\cdots\le e_m$ those of $h_{N,S}$. The eigenvalues of $\rho_{N,S}$ are $q_a/(2S+1)$, each repeated
$2S+1$ times, whereas the energies are $e_a$ with the same
multiplicity. Their passive pairing gives $\sum_{a=1}^m q_ae_a$. This minimum
is attained by a unitary of the form
$u_{N,S}\otimes\mathbb I_{2S+1}$, which commutes with
$S_x$, $S_y$, and $S_z$. Summing over the joint sectors
therefore establishes, for the present spin-invariant states,
\begin{equation}
\begin{aligned}
\mathcal W_{U(1)\times SU(2)}(\rho)
&=\WNS(\rho),\\
\mathcal L^{\mathrm{mix}}_{U(1)\times SU(2)}(t)
&=\LNS(t),
\end{aligned}
\label{eq:SU2_equivalence}
\end{equation}
where the first optimization is over unitaries commuting
with both $\Ne$ and all global spin components. These
equalities hold for $\rho=\rho_0$, $\rho_r(t)$, and
$\bar\rho(t)$ in the stated protocol. They do not establish
that the electrical control operators $\Tb$ and $\ZD$
generate every admissible $SU(2)$-invariant unitary.

\section{Exact spectral formula for the singlet--triplet work deficit}
\label{app:ST_closed_form}
We prove Eq.~\eqref{eq:exact_ST_gap} for a spin-rotation-invariant
conditional $N=2$ state in the repulsive regime $U>V$.
Write $a=E_-^{(2)}$, $d=E_+^{(2)}$, and recall the ordered six-level energy list
$(a,V,V,V,U,d)$, with $a<V<U<d$. Spin conservation permits independent passive
rearrangement of the singlet and triplet blocks, giving
\begin{equation}
 E_{\mathrm{pass},2}^{N,S}=a s_1+U s_2+d s_3+3Vq.
 \label{eq:ST_pass_energy}
\end{equation}
By contrast, the $U(1)$-constrained passive energy results from
sorting the multiset $\{s_1,s_2,s_3,q,q,q\}$ against the six
ordered energies. There are exactly four distinct orderings of the
triplet weight $q$ relative to the singlet eigenvalues.
For compactness let $\mathscr P_U=E_{\mathrm{pass},2}^{U(1)}$:
\begin{equation}
\begin{aligned}
q\le s_3:\quad
 &\mathscr P_U
 =a s_1+V(s_2+s_3+q)+Uq+dq,\\
s_3\le q\le s_2:\quad
 &\mathscr P_U
 =a s_1+V(s_2+2q)+Uq+ds_3,\\
s_2\le q\le s_1:\quad
 &\mathscr P_U
 =a s_1+3Vq+Us_2+ds_3,\\
s_1\le q:\quad
 &\mathscr P_U
 =a q+V(2q+s_1)+Us_2+ds_3.
\end{aligned}
\label{eq:ST_four_orderings}
\end{equation}
Subtracting each expression from Eq.~\eqref{eq:ST_pass_energy}
gives the four respective work deficits per unit $N=2$ weight:
\begin{equation}
\begin{cases}
(U-V)(s_2-q)+(d-V)(s_3-q),&q\le s_3,\\
(U-V)(s_2-q),&s_3\le q\le s_2,\\
0,&s_2\le q\le s_1,\\
(V-a)(q-s_1),&s_1\le q.
\end{cases}
\label{eq:ST_four_gaps}
\end{equation}
The branches join continuously at the boundaries. They combine
exactly into the positive-part expression of
Eq.~\eqref{eq:exact_ST_gap}. As $p_2>0$, $U>V$, $t_0>0$ imply strictly
positive energy gaps, the deficit is zero exactly when
$s_2\le q\le s_1$. The term $(V-a)(q-s_1)_+$ is needed for a
general spin-invariant state with unusually large triplet
occupation; it vanishes for the reference noise-averaged state.
The equality with the difference of the two mixing losses further
requires the initial state to be passive under both constraints, as
is the grand-canonical thermal preparation in this work. This
specific threshold is not asserted for spin-anisotropic dynamics
or Hamiltonians that split the three triplet energies.

\section{Weak-noise activation of the singlet--triplet work deficit}
\label{app:noise_onset}
\subsection{Driven parameter-response generator}
Work in the conditional $N=2$ singlet manifold and dimensionless
time $\tau=t t_0/\hbar$. Write its Hamiltonian in units of $t_0$
as $h_\eta(\tau)=h_0(\tau)+\eta Z_{D,2}$, where
$\eta=\delta\epsilon/t_0$ and $h_0(\tau)$ includes the exact resonant barrier modulation
and the envelope of Eq.~\eqref{eq:drive_envelope_numeric}.
Its propagator satisfies
$\partial_\tau U_\eta=-i h_\eta U_\eta$ and $U_\eta(0)=I$.
Differentiating at $\eta=0$ gives the Hermitian response generator
\begin{equation}
 K_\tau=iU_0^\dagger(\tau)
 \left.\frac{\partial U_\eta(\tau)}{\partial\eta}\right|_{\eta=0}
 =\int_0^\tau\!ds\,U_0^\dagger(s)Z_{D,2}U_0(s).
 \label{eq:app_response_generator}
\end{equation}
This is the standard generator for a unitary dependence on a parameter
\cite{PangBrun2014}. In particular,
$U_\eta(\tau)=U_0(\tau)[I-i\eta K_\tau+O(\eta^2)]$.

\subsection{Perturbation of the almost rank-one thermal state}
For the low-temperature reference preparation, the conditional
singlet block before charging has the form
$\rho_s(0)=\lambda_g|g\rangle\langle g|+R_{\rm exc}$,
with $\|R_{\rm exc}\|_1=\lambda_{\rm exc}\ll1$.
The thermal triplet density matrix is $qI_3$, and spin-independent
unitary evolution leaves $q$ unchanged.
For a centered Gaussian $\eta$ of width
$\sigma=\sigma_\epsilon/t_0$, define the interaction-picture average
\begin{equation}
 \widetilde\rho_s(\sigma,\tau)=
 \mathbb E_\eta\left[U_0^\dagger(\tau)U_\eta(\tau)
 \rho_s(0)U_\eta^\dagger(\tau)U_0(\tau)\right].
 \label{eq:app_ensemble_interaction}
\end{equation}
This conjugation leaves all density-matrix eigenvalues unchanged.
Let $P_\perp=I-|g\rangle\langle g|$ project onto the
\emph{two-dimensional} excited-singlet subspace. Expanding the
random pure-state projector, averaging the centered distribution,
and projecting on both sides yields
\begin{equation}
\begin{aligned}
 P_\perp\widetilde\rho_s P_\perp
 &=\lambda_g\sigma^2\,
 P_\perp K_\tau|g\rangle\langle g|K_\tau P_\perp\\
 &\quad+O(\sigma^4)+O(\lambda_{\rm exc}).
\end{aligned}
 \label{eq:app_excited_spectral_projection}
\end{equation}
The projected leading term is positive and of rank at most one.
Its only possibly nonzero eigenvalue is
\begin{equation}
 \chi(\tau)\sigma^2,
 \qquad \chi(\tau)=\lambda_g\|P_\perp K_\tau|g\rangle\|^2.
 \label{eq:app_chi}
\end{equation}
Terms of the form $|g\rangle\langle\cdot|$ generated by the
second derivative of $U_\eta$ vanish under double projection.
Diagonalizing \emph{within} the excited-singlet subspace, rather
than applying separate nondegenerate perturbation formulas to its
nearly vanishing thermal eigenvalues, gives
\begin{equation}
 s_2(\sigma,\tau)=\chi(\tau)\sigma^2
 +O(\sigma^4)+O(\lambda_{\rm exc}).
 \label{eq:app_s2}
\end{equation}
The spectral perturbation procedure follows standard
finite-dimensional methods \cite{Kato1995}; the specific
coefficient comes from the driven Hubbard singlet Hamiltonian.
For fixed finite time this expansion is controlled as
$\sigma\to0$; the separate thermal remainder need not be
small relative to $\sigma^2$ arbitrarily close to $\sigma=0$.

\subsection{Threshold and numerical verification}
The exact work-deficit formula~Eq.~\eqref{eq:exact_ST_gap} implies that its first appearance on
the physically relevant branch $q<s_1$ occurs at $s_2=q$.
Combining this with Eq.~\eqref{eq:app_s2} yields the leading
prediction Eq.~\eqref{eq:predicted_sigma_threshold}, provided
$\chi(\tau)>0$ and the crossing remains in the perturbative regime.
At fixed drive and time, the low-temperature approximation
$q\simeq\exp[-(V-E_-^{(2)})/(k_BT)]$
then suggests the leading scaling
$\sigma_{\epsilon,c}/t_0\sim\chi(\tau)^{-1/2}
\exp[-(V-E_-^{(2)})/(2k_BT)]$, with the understood
weak temperature dependence of $\chi$. This estimate does not
establish a finite positive threshold at $T=0$.

For $U/t_0=14$, $V/t_0=3$, $k_BT/t_0=0.05$,
$A_t/t_0=0.10$, and $\tau=17$, the same envelope and resonant
frequency as in the main figures give
$\chi(17)\simeq11.7557$ and $q\simeq0.000867751$.
The leading threshold is $0.008592$, compared with a
numerical first crossing bracketed by $0.00864$ and $0.00865$.
The nonlinear dynamics are propagated with unitary midpoint
steps of $\delta\tau=0.005$; the Gaussian average uses 13-point
probabilists' Gauss--Hermite quadrature (checked against higher
order at representative points). The numerical test directly
compares the two spectral quantities, $s_2-q$, rather than fitting
a kink in an ergotropy curve.
\begin{table}[!ht]
\caption{\textbf{Numerical bracket for the first noise-induced spin-work-deficit onset.}
The spectral difference $s_2-q$ changes sign between
$\sigma_\epsilon/t_0=0.00864$ and $0.00865$ at $\tau=17$.
Parameters: $U/t_0=14$, $V/t_0=3$, $A_t/t_0=0.10$,
$k_BT/t_0=0.05$, $\mu=\mu_{12}^{(T)}$;
unitary propagation uses $\delta\tau=0.005$ and
13-point Gauss--Hermite averaging. Both tabulated quantities
are dimensionless.}
\label{tab:threshold_bracket}
\begin{ruledtabular}
\begin{tabular}{rc}
$\sigma_\epsilon/t_0$ & $s_2-q$\\ \hline
0.00600 & $-4.46983\times10^{-4}$\\
0.00864 & $-6.13168\times10^{-7}$\\
0.00865 & $+1.37133\times10^{-6}$\\
0.00900 & $+7.22032\times10^{-5}$
\end{tabular}
\end{ruledtabular}
\end{table}
The numerical verification procedure checks both the onset and the
four-branch spectral formula against direct eigenvalue sorting;
it does not claim convergence of every curve in the manuscript.

\section{Thermal factorization and crossover checks}
\label{app:thermal_onsets}
The initial singlet block, with its normalization inherited from
the conditional $N=2$ state, decomposes as
\begin{equation}
 \rho_s(0;T)=\lambda_g(T)|g\rangle\langle g|+R_{\rm exc}(T),
 \qquad R_{\rm exc}(T)\succeq0,
 \label{eq:exc_singlet_decomp}
\end{equation}
with
\begin{equation}
 \lambda_{\rm exc}(T)=\Tr R_{\rm exc}(T)
 =\frac{e^{-\beta U}+e^{-\beta E_+^{(2)}}}{Z_2}.
 \label{eq:exc_singlet_weight}
\end{equation}
Linearity and positivity of the same singlet random-unitary channel
then give
\begin{equation}
 \bar\rho_s(T)=\lambda_g(T)A_g+R_T,\quad R_T\succeq0,
 \quad\Tr R_T=\lambda_{\rm exc}(T),
 \label{eq:exc_averaged_decomp}
\end{equation}
where the dependence on fixed $\sigma_\epsilon$ and $\tau$ is
suppressed. Since $0\preceq R_T\preceq\lambda_{\rm exc}I_3$,
the min--max principle for ordered Hermitian eigenvalues yields
\begin{equation}
 0\le\frac{s_i}{\lambda_g}-a_i\le\delta_{\rm th}(T),\qquad
 i=1,2,3,
 \label{eq:thermal_error_bound}
\end{equation}
with
\begin{equation}
 \delta_{\rm th}(T)=\frac{\lambda_{\rm exc}(T)}{\lambda_g(T)}
 =e^{-(U-E_-^{(2)})/(k_BT)}+e^{-\Delta_{\rm eff}/(k_BT)}.
 \label{eq:thermal_delta_bound}
\end{equation}
This bound remains valid at eigenvalue crossings and is uniform in
charging time and noise strength for a temperature-independent channel.
For $U/t_0=14$ and $V/t_0=3$, the gaps are
$(U-E_-^{(2)})/t_0=(11+\sqrt{137})/2$ and
$\Delta_{\rm eff}/t_0=\sqrt{137}$, whence
$\delta_{\rm th}(T)<5\times10^{-17}$ throughout
$0<k_BT/t_0\le0.30$. This controls the omitted thermal
contribution to scaled singlet eigenvalues; it does not bound
numerical propagation, Gaussian-quadrature, or crossover-temperature
errors. Equation~\eqref{eq:semi_analytic_Tc} uses the fully averaged
$A_g$ without extending the weak-noise expansion beyond its domain.

The temperature at which the second singlet eigenvalue crosses an
individual triplet weight is located numerically from
$s_2(T,\sigma_\epsilon;\tau)-q(T)=0$ at $\tau=17$. The results
below use the same loading convention $\mu=\mu_{12}^{(T)}$ at every
preparation temperature. In the higher-noise region the Gaussian quadrature
order is increased to test the stability of the crossing.
\begin{table}[!ht]
\caption{\textbf{Thermal singlet--triplet crossover temperatures for quasistatic detuning noise.}
At each noise amplitude $\sigma_\epsilon/t_0$, the boundary
$k_BT_c/t_0$ is located by $s_2(T_c)=q(T_c)$ on the branch $q<s_1$,
with $\tau=17$, $U/t_0=14$, $V/t_0=3$,
$A_t/t_0=0.10$, and $\mu=\mu_{12}^{(T)}$.
The values are rounded to the precision supported by the reported
propagation and Gaussian-quadrature refinements.}
\label{tab:thermal_noise_thresholds}
\begin{ruledtabular}
\begin{tabular}{cc}
$\sigma_\epsilon/t_0$ & $k_BT_c/t_0$\\
\hline
0.030 & 0.07525\\
0.060 & 0.09740\\
0.100 & 0.11607\\
0.150 & 0.13458\\
0.200 & 0.17827\\
0.250 & 0.22011\\
0.300 & 0.25468
\end{tabular}
\end{ruledtabular}
\end{table}

\section{Parity-resolved crossover and independent convergence checks}
\label{app:parity_convergence}
Equation~\eqref{eq:averaged_parity} can be checked directly by
computing $\|\mathcal P_S\bar\rho_s\mathcal P_S^\dagger-
\bar\rho_s\|$, which is of order $10^{-17}$ for the symmetric
Gaussian ensembles tested. For consistency with Table~\ref{tab:thermal_noise_thresholds}, the
parity-resolved entries below use $\delta\tau=0.005$ and $n_q=41$.
The internal eigenvalue difference
$\delta_{\rm parity}=\lambda_{{\rm even},-}-\lambda_{\rm odd}$
changes sign between $\sigma_\epsilon/t_0=0.145$ and $0.150$
when evaluated \emph{on} the boundary $s_2=q$. The corresponding
boundary temperatures and parity differences are reported in
Table~\ref{tab:parity_branch_crossing}.
\begin{table}[!ht]
\caption{\textbf{Parity-resolved singlet eigenvalue reordering on the thermal crossover boundary.}
At each $\sigma_\epsilon/t_0$, the temperature $T_c$ satisfies $s_2=q$;
$\delta_{\rm parity}=\lambda_{{\rm even},-}-\lambda_{\rm odd}$
is the difference between the smaller even-parity singlet eigenvalue
and the odd-parity eigenvalue.
Parameters: $U/t_0=14$, $V/t_0=3$, $A_t/t_0=0.10$,
$\tau=17$, $\delta\tau=0.005$, $n_q=41$, and 
$\mu=\mu_{12}^{(T)}$.}
\label{tab:parity_branch_crossing}
\begin{ruledtabular}
\begin{tabular}{ccc}
$\sigma_\epsilon/t_0$ & $k_BT_c/t_0$ & $\delta_{\rm parity}$\\
$0.140$ & $0.129331$ & $-0.003689$\\
$0.145$ & $0.130628$ & $-0.000354$\\
$0.150$ & $0.134579$ & $+0.003168$\\
\end{tabular}
\end{ruledtabular}
\end{table}
The parity crossing involves the eigenvalues of the averaged singlet density matrix, not the
energy levels of the Hubbard Hamiltonian. It is distinct from the condition $s_2=q$ defining
the additional spin-work crossover.
To test numerical stability, we located $s_2-q=0$ by root finding
for each independently fixed time step $\delta\tau$ and
probabilists' Gauss--Hermite order $n_q$. Propagation uses the same
midpoint unitary algorithm and pulse envelope as the main figures.
\begin{table}[!ht]
\caption{\textbf{Separate time-step and quadrature refinements of the thermal crossover.}
The boundary $s_2(T_c)=q(T_c)$ is located at $\tau=17$ for
$\sigma_\epsilon/t_0=0.10$, $0.30$, and $0.35$, with $U/t_0=14$,
$V/t_0=3$, $A_t/t_0=0.10$, and $\mu=\mu_{12}^{(T)}$.
The columns vary the scaled propagation step $\delta\tau$ and
Gauss--Hermite order $n_q$ independently; extra digits in
$k_BT_c/t_0$ are shown only to expose numerical differences.}
\label{tab:separate_crossover_convergence}
\begin{ruledtabular}
\begin{tabular}{cccc}
$\sigma_\epsilon/t_0$ & $\delta\tau$ & $n_q$ & $k_BT_c/t_0$\\
\hline
0.10 & 0.010 & 25 & 0.11606578\\
0.10 & 0.010 & 41 & 0.11606578\\
0.10 & 0.005 & 25 & 0.11607284\\
0.10 & 0.005 & 41 & 0.11607284\\
0.30 & 0.010 & 25 & 0.25602750\\
0.30 & 0.010 & 41 & 0.25471343\\
0.30 & 0.010 & 61 & 0.25471007\\
0.30 & 0.005 & 25 & 0.25599439\\
0.30 & 0.005 & 41 & 0.25468275\\
0.30 & 0.005 & 61 & 0.25467939\\
0.35 & 0.005 & 61 & 0.27790009\\
0.35 & 0.005 & 81 & 0.27790004\\
0.35 & 0.0025 & 61 & 0.27789474\\
0.35 & 0.0025 & 81 & 0.27789469\\
\end{tabular}
\end{ruledtabular}
\end{table}
At $\sigma_\epsilon/t_0=0.30$, the 25-point quadrature is
insufficient even if the time step is halved. The 41-to-61-point
change at $\delta\tau=0.005$ is approximately
$3.36\times10^{-6}$ in dimensionless crossover temperature;
this is numerically small compared with the boundary scale but
larger than the precision of an unqualified eight-digit result.
At $\sigma_\epsilon/t_0=0.35$, 61 and 81 quadrature points give
nearly identical results at fixed time step, while halving $\delta\tau$
changes $k_BT_c/t_0$ by approximately $5.4\times10^{-6}$.
This additional test covers the largest noise amplitude shown in
Fig.~\ref{fig:thermal_noise_crossover}(a). The values quoted in the main
text are rounded to the stability supported by these checks;
no certified uniform error bound is asserted for every point in the map.

\section{Thermally weighted barrier response (auxiliary diagnostic)}
\label{app:barrier_response}
We can also assess the transition strengths through the barrier response weighted by the thermal populations.
For the eigenstates of $H_B$ with equilibrium populations $w_n^{\rm th}$,
we define the positive-frequency spectral weight by
\begin{equation}
\begin{aligned}
\mathcal A_T(\omega)={}&\sum_{m>n}(w_n^{\rm th}-w_m^{\rm th})
|\langle E_m|\Tb|E_n\rangle|^2 \\
&\times\delta_\gamma[\hbar\omega-(E_m-E_n)].
\end{aligned}
\label{eq:thermal_response}
\end{equation}
The eigenstates are ordered by increasing energy, and
$w_n^{\rm th}=\langle E_n|\rho_{\beta,\mu}|E_n\rangle$.
The symbol $\delta_\gamma$ denotes a normalized broadened peak centered
at zero; its precise profile does not affect the integrated line weight.
At zero detuning, the barrier operator connects only the two even
singlets at positive frequency. Therefore Eq.~\eqref{eq:thermal_response}
reduces exactly to
\begin{equation}
\mathcal A_T(\omega)=p_2 M_{\rm tun}
\frac{e^{-\beta E_-^{(2)}}-e^{-\beta E_+^{(2)}}}{Z_2}
\,\delta_\gamma(\hbar\omega-\Delta_{\rm eff}).
\label{eq:thermal_response_exact}
\end{equation}
This expression excludes the finite but barrier-dark
$N=1$ and $N=3$ splittings and the odd doublon. It also
vanishes when $U=V$, in agreement with Eq.~\eqref{eq:Mtun}.
No broadened spectrum is used in the numerical simulations;
the main text compares the exact transition matrix element and
gap curvature in Fig.~\ref{fig:interaction_selectivity}.

\section*{Data Availability}
The numerical data and simulation scripts supporting the reported results
are available from the corresponding author upon reasonable request.
\clearpage
\bibliography{khalil_clean}

\end{document}